 \documentclass[final,3p,times,twocolumn]{elsarticle}
\usepackage{amssymb}
\usepackage{amsmath}
\usepackage{amsthm}
\usepackage{algorithm}
\usepackage{eurosym}
\usepackage{algpseudocode}
\newtheorem{definition}{Definition}
\newtheorem{theorem}{Theorem}
\newtheorem{nono-theorem}{Theorem}[]
\makeatletter
\newenvironment{heuristic}[1][htb]{%
    \renewcommand{\ALG@name}{Heuristic}% Update algorithm name
    \begin{algorithm}[#1]%
    }{\end{algorithm}
}
\makeatother

\newcommand{\LPVY}{LPVY$(z, M_p^j, b_p^j)$}
\newcommand{\LPvm}{LP-vm$(z, M_p^j, b_p^j)$}
\newcommand{\LPvM}{LP-vM$(z, M_p^j, b_p^j)$}
\newcommand{\LPym}{LP-ym$(z, M_p^j, b_p^j)$}
\newcommand{\LPyM}{LP-yM$(z, M_p^j, b_p^j)$}

\newcommand\rsetminus{\mathbin{\mathpalette\rsetminusaux\relax}}
\newcommand\rsetminusaux[2]{\mspace{-4mu}
  \raisebox{\rsmraise{#1}\depth}{\rotatebox[origin=c]{-20}{$#1\smallsetminus$}}
 \mspace{-4mu}
}
\newcommand\rsmraise[1]{%
  \ifx#1\displaystyle .8\else
    \ifx#1\textstyle .8\else
      \ifx#1\scriptstyle .6\else
        .45%
      \fi
    \fi
  \fi}

\DeclareMathOperator{\cb}{cne}

\journal{Energy Economics}

\begin{document}

\begin{frontmatter}



\title{On Market Clearing of Day Ahead Auctions for European Power Markets: Cost Minimisation versus Social Welfare Maximisation}


\author[inst1, inst2]{Ioan Alexandru Puiu}

\affiliation[inst1]{organization={Mathematical Institute, University of Oxford},%Department and Organization
            addressline={Andrew Wiles Building, Woodstock Rd}, 
            city={Oxford},
            postcode={OX26GG}, 
            state={Oxfordshire},
            country={United Kingdom}}

\author[inst1]{Raphael Andreas Hauser}

\affiliation[inst2]{organization={Corresponding author, email: ioan.puiu@maths.ox.ac.uk}%Department and Organization
            }

\begin{abstract}
For the case of inflexible demand and considering network constraints, we introduce a Cost Minimisation (CM) based market clearing mechanism, and a model representing the standard Social Welfare Maximisation mechanism used in European Day Ahead Electricity Markets. Since the CM model corresponds to a more challenging optimisation problem, we propose four numerical algorithms that leverage the problem structure, each with different trade-offs between computational cost and convergence guarantees. These algorithms are evaluated on synthetic data to provide some intuition of their performance.  We also provide strong (but partial) analytical results to facilitate efficient solution of the CM problem, which call for the introduction of a new concept: optimal zonal stack curves, and these results are used to devise one of the four solution algorithms. An evaluation of the CM and SWM models and their comparison is performed, under the assumption of truthful bidding, on the real world data of Central Western European Day Ahead Power Market during the period of 2019-2020. We show that the SWM model we introduce gives a good representation of the historical time series of the real prices. Further, the CM reduces the market power of producers, as generally this results in decreased zonal prices and always decreases the total cost of electricity procurement when compared to the currently employed SWM. 
\end{abstract}

% \centering
% \includegraphics[width = 16cm]{P3Gabstract.pdf}

% \begin{highlights}
% \item We propose a Cost Minimisation and a Social Welfare Maximisation market clearing model with network constraints
% \item We obtain strong (but partial) analytical results for the Cost Minimisation model
% \item We devise numerical algorithms to solve the more challenging Cost Minimisation problem
% \item Our SWM model represents well the historical price data of the Central Western European spot market
% \item The proposed Cost Minimisation approach generally results in reducing clearing prices
% \end{highlights}

%\centering
%\includegraphics[width = 16cm]{Paper 3. On market %clearing - graphical abstract.pdf}

%\item We propose a Cost Minimisation and a Social %Welfare Maximisation market clearing model with network %constraints
%\item We obtain strong (but partial) analytical results %for the Cost Minimisation model
%\item We devise numerical algorithms to solve the more %challenging Cost Minimisation problem
%\item Our SWM model represents well the historical %price data of the Central Western European spot market
%\item The proposed Cost Minimisation approach generally %results in reducing clearing prices


\end{frontmatter}
\section{Introduction}\label{sec:Intro}
Electricity spot markets have undertaken heavy deregulation in the past two decades, and the goal of this liberalisation is to improve economic efficiency and attract new investments \citep{Wilson2002}. However, for this to be achieved, each electricity market requires a market clearing mechanism to facilitate the desired outcome. According to the work of \citet{Myersonimpossib} there are four main desirable properties that a market outcome needs to satisfy, which can be informally be stated as: (i) each market participant has an incentive to participate, (ii) consumers' payment covers producers' revenue (iii) truthful bidding is incentivised, and (iv) the good is given to the one who values it most. However, according to the \textit{Myerson-Satterthwaite impossibility theorem} \cite{Myersonimpossib}, for the case of two agents and one good, the four properties mentioned above cannot be simultaneously satisfied, suggesting that any mechanism must sacrifice at least one. This result, combined with the difficulty of electricity storage and transmission, and different preferences expressed by different policy makers, are perhaps the main reason for which a plethora of market clearing mechanisms are used throughout the deregulated electricity markets' world. These mechanisms can have very different rules and properties and thus the theoretical, numerical or experimental results on one market cannot be simply generalised to another, at least not without careful consideration. Nevertheless, a thorough understanding of each particular market is of great interest for the corresponding policy makers, market participants, and researchers. The importance of understanding electricity spot markets is also extended to market participants that do not trade on the spot market. This is because futures or forward contracts are strongly influenced by the expectation of future spot prices, while options are normally written on the futures contracts prices \cite{DavisBook}. Thus, electricity spot markets lie the foundation of all electricity markets.
\subsection{Brief literature survey and motivation}
In this paper we focus on understanding, modelling and formulating numerical solutions to market clearing mechanisms consistent with the policy requirements of European countries. In particular, the Central Western European (CWE) Day Ahead Auction (DAA) is considered as a case study. Since the Day Ahead Market represents the largest part of the spot electricity markets, we abuse nomenclature and use these two terms interchangeably here. To the best of our knowledge, when compared to other markets, and their corresponding clearing mechanism, the CWE spot market is not very well studied from either a theoretical or numerical simulations point of view. The CWE DAA market uses Flow-Based Market Coupling (FBMC) grid constraints, and zonal marginal pricing, which are very different compared to their more widely used counterparts of Optimal Power Flow (OPF) based constraints, and Locational Marginal Pricing (LMP). For detailed reviews of how FBMC grid constraints are obtained we refer the reader to \citep{WPEN2014}, \citep{WPEN2015}, while the official documentation can be found at \citep{JAO}, but we briefly introduce the required concepts in Sub-section \ref{sec:background}.
\par 
The work done by the research community on market clearing mechanisms can generally be split in two: (i) auction solution, concerned with building, solving and applying models to real world and (ii) auction design, generally concerned with defining and analysing market clearing mechanisms in the search for achieving desirable properties. This distinction is usually, but not always very obvious. Our work is placed at the intersection of auction design and auction solution, with much stronger emphasis on the latter. We are concerned with modelling the current market clearing mechanism for the CWE, which is based on a Social Welfare Maximisation (SWM) objective, subject to supply and demand balance, grid constraints, and producers' capacity constraints, while considering demand as inflexible. We also propose a Cost Minimisation objective as an alternative for reducing producer's market power, which is an usual concern in electricity markets \cite{HowisonStack}.  The Cost Minimisation objective is not very popular in either theory or practice, perhaps due to the mathematical complexity added by requiring explicit modelling of price outcome variables in the market clearing problem. We here provide efficient numerical solutions for our CM market clearing and show the potential benefits of this mechanisms.
\par 
Some of the most relevant research on the auction solution front is the work of \citet{HowisonStack}, which build a fundamental stack model where various underlying factors are modelled as stochastic processes. The model completely ignores grid structure, considers demand as inelastic, requires availability of order book data, and only models daily peak prices (excluding intra-day variations). However, this model cannot be directly applied to the CWE, as the order book data is not available for this market, and network constraints are very important for price formation. In addition, estimating intra-day variations may be very desirable. Further, as we shall see later, a stack model is simply the analytic solution (when such solution is possible) of a particular case of cost minimisation market clearing. Cost Minimisation objective for spot electricity markets is also considered in \citet{Alguacil2014}, which requires explicit modelling of outcome prices as decision variables. An Optimal Power Flow grid model combined with Locational Marginal Pricing are used (these are popular choices in North American markets). The resulting optimisation problem is bi-level mixed integer linear program and is generally very computationally intense. While this model is very detailed and a numerical solution approach is provided, the model cannot be directly applied to the CWE market due to the OPF and LMP choices. Further, no results are provided for real world markets to evaluate the suitability of the model. The related work of \citet{Alguacil2013} focuses on price uniqueness issues for OPF grid modelling with LMP pricing. A similar OPF-LMP model was considered under a Game-Theoretic framework in \citep{DRalph2007}, but the orders are parametrised as linear marginal prices with respect to quantity. 
\par Auction Mechanism Design has received extensive academic attention, with the end goal of obtaining desirable auction properties, but these studies and proposed mechanisms are not generally applicable to the CWE spot market for various reasons. Some of the most recent work is focused on a second-price type of auction mechanism, known as the Vickrey-Clarke-Groves (VCG) which has remarkable theoretical properties \cite{ETH2016}, \citep{ETH2017}. However, VCG is unpopular in practice, as the desirable properties come at increased computational cost, and the violation of weak balance of payments criterion, meaning that auctioneers have to pay for achieving the other desirable properties. Further, the mechanism is not collusion proof, and as a results of this a \textit{core-selecting} (CS) mechanism is proposed in \citep{OrcunCore}. However, the core selecting mechanism is even more computationally intense, and also does not satisfy weak balance of payments. The desirable properties of VCG and CS mechanisms are only proved for the case where prices do not appear as explicit variables in the market clearing optimisation problem. For these reasons, and the zonal pricing requirement in the CWE, neither of the two mechanisms can be used directly. 
\subsection{Main contributions}
In contrast to \cite{HowisonStack}, \citep{Alguacil2013}, \cite{Alguacil2014} and most of the work on auction solution in the literature, we build a market clearing model consistent with the CWE methodology \cite{JAO}, show how to solve this model efficiently, and apply it to real world data, showing its ability to represent actual prices time series well. Our model considers specific features such as: Social Welfare Maximisation objective, FBMC network constraints and zonal marginal pricing (ZMP). Similarly to \citep{DRalph2007}, the orders are parametrised as linear marginal prices with respect to quantities. As suggested by \citep{HowisonStack}, the demand side is generally inflexible, and thus we consider the case of purely inflexible demand. To allow us to focus on the auction mechanism and its solution, we assume that the orders are fully defined by the producers' cost. This is also suggested in the work of \citet{HowisonStack}, and is a commonly used assumption in the CWE spot market. We also investigate applying a Cost Minimisation objective to the CWE market clearing, solution approaches, and model's performance on real data corresponding to the CWE. To our knowledge this was not previously studied. The case of (unique) zonal marginal prices is in stark constrat with locational marginal prices since unlike for the latter, when demand is taken as inflexible, the SWM objective \textit{does not} reduce to a CM objective in the ZMP case. Instead, SWM objective reduces to some form of \textit{"apparent cost"} which may lead to very different market outcomes. 
\par
To the best of our knowledge, there is scarcity of auction design and auction solution research work directly applicable to the European spot electricity markets. This is particularly important as the FBMC methodology is gaining momentum, with 7 more countries being incorporated in 2022, versus the 5 countries initially included in 2014.
The main contributions of our work are an attempt to expand the knowledge on the European spot electricity markets, and can be summarised as following:
\begin{enumerate}
    \item We provide a detailed SWM-based market clearing model to represent the specific features of the CWE market, while retaining computational tractability, and we show how this can be solved for efficiently. 
    \item We propose a Cost Minimisation based market clearing, that requires the inclusion of zonal prices as explicit variables in the objective and constraints. 
    \item We devise four numerical approaches to solve for the CM market clearing model, each with different trade-offs between computational cost and convergence guarantees. Either of these approaches is much less computationally intense than the somewhat equivalent Mixed Integer Linear Program counterparts proposed in \citep{Alguacil2013}.
    \item We obtain strong (although partial) analytical results for the CM model: we are able to analytically compute what we call \textit{zonal stack curves} that are then used by one of our four numerical algorithms to efficiently obtain a solution to the CM, with stronger than local convergence guarantees.
    \item We apply both the SWM and the CM based models to the real world data of CWE for the period of 2019-2020, solving on an hourly basis for 5016 hours in total. To our knowledge, this type of results for fundamental models considering network and capacity constraints, demand and impact of fuel prices are unprecedented, and confirm the computational tractability of our numerical approaches.
    \item The results obtained by the SWM model represent the historical time series of prices well. We also show that the CM model reduces producers' market power by generally resulting in reduced market prices, and always reducing the total procurement cost.
\end{enumerate}
The rest of this section is dedicated to the introduction of basic pre-requisite concepts, after which we conclude the section by reviewing the structure of our paper.
\subsection{Background information}\label{sec:background}
We here briefly review what is meant by an auction mechanism, the model of the network constraints in the CWE, a simplified representation of the CWE market clearing mechanism and finally, the case of inflexible demand.
\par Unlike spot financial markets, spot electricity markets are usually split in two, due to the difficulty of storing large amounts of electricity, and uncertainty in demand and production patterns. The vast majority of spot trading and capacity allocation occurs during Day Ahead Auctions (which are resolved one day prior the delivery time) while the Intra-day balancing markets ensure balancing of supply and demand under feasibility constraints. Spot financial markets usually allow for two types of orders: (i) limit orders, that only clear if the set price is offered or asked by a counter-party, and (ii) market orders that are resolved at the best available counter-party limit order price. Trading occurs continuously and price changes so long as a new trading agreement is struck. By contrast, Day Ahead electricity markets only effectively allow for limit orders, since they are cleared just once, prior to the delivery, for a specified time interval which is usually 15 minutes or 1 hour. This is to ensure supply and demand are balanced simultaneously with other market specific feasibility constraints, such as for example feasibility of transmission. 
\par 
Any such market clearing mechanism then has an objective, which for example could be Social Welfare Maximisaiton or Cost Minimisation, and constraints such as supply and demand balance, network transmission constraints, producers' capacity constraints and ramping constraints \cite{mihaRH2015} and others. This means that any market clearing problem can be expressed as a mathematical optimisation problem. The mechanism must include a quantity allocation and a pricing mechanism. Often, to maintain mathematical simplicity, quantity allocation is performed first, after which the price is determined \citep{JAO}, \citep{ETH2017}.
\par
Let us now review a simplified version of the market clearing mechanism used in the CWE Day Ahead Auction. The market clearing problem in CWE DAA has integrality constraints corresponding to fill-or-kill
orders or blocks of orders. However, we here neglect these constraints for simplicity and focus on the pricing mechanism when any fraction of a bid or offer could be accepted. Supply and demand balance, as well as network constraints via the Flow Based Market Coupling (FBMC) methodology are also considered \citep{JAO}. The main difference between FBMC constraints and OPF ones is that the FBMC approach makes approximations that enable treating every node within a node cluster, called a zone (usually a country), as equivalent, while OPF considers the impact of every nodal power change on every line. With the current CWE DAA Market clearing mechanism, the market outcome is determined as following:
\begin{enumerate}
    \item Solving a (network constrained) primal problem corresponding to the Social Welfare Maximisation, here modelled as an LP.
\item Solving the corresponding dual problem to obtain the Lagrangian multipliers of the network constraints,
\item  and solving a pricing problem accounting for the marginal transmission costs, and a somewhat arbitrary objective that ensures that the price is ”intuitive”.
\end{enumerate}
The objective of the primal LP is maximisation of the social welfare expressed as
\begin{equation}
    SW(x_.) = \sum_{z \in \mathcal{Z}}\left( \sum_{b \in \mathcal{B}_z}Q_b^zP_b^zx_b^z  - \sum_{b \in \mathcal{O}_z}Q_o^zP_o^zx_o^z\right),
\end{equation}
where $\mathcal{Z}$ is the set of zones and $\mathcal{B}_z$ and $\mathcal{O}_z$ are the sets of bids and offers for zone $z$, and $x_.$ is the vertical concatenation of all the accepted quantity fractions variables $x^z \in [0,1]$. The quantities $(Q_b^z, P_b^z)$ and $(Q_o^z, P_o^z)$, are non-negative constants corresponding to quantity-price pairs for the bids and offers respectively.
\par 
Network constraints are represented by the combination of line-zonal Power Transmission Distribution Factors (PTDFs), and lines' Remaining Available Margins (RAMs). The PTDFs measure the impact of one unit of zonal power change on the transmission lines, while the two RAMs corresponding to each line quantify the maximum feasible change limits on this line. A detailed review of how these constraints are obtained is provided in \citep{WPEN2014}, but what matters here is that the resulting network constraints are linear, and applied at net zonal power levels. Available historical data of these constraints is very sparse \citep{JAOTool}, \citep{PuiuHauser2021}, but a methodology to obtain an estimate of the full set of constraints is given in \citep{PuiuHauser2021}. We use these results as inputs for our case study.
\par 
Let $E \in \mathbb{R}^{|\mathcal{Z}| \times N_b}$ be a linear map from the accepted orders fractions to the net energy production in the zone. The entries corresponding to offers are assigned $-Q_o^z$ values since positive values are associated with consumption. The CWE network constraints can then be represented as :
\begin{equation}
    M_p \hat{y} \leq b_p,
\end{equation}
where $\hat{y}=Ex$ is the vector of net zonal productions, $M_p^+$ is the concatenation of the line-zonal PTDFs corresponding to time $t$ and all considered scenarios $s$ and $M_p= ||_v(-M_p^+,M_p^+)$. A similar concatenation over the RAMs is performed: $b_p= ||_v(b_p^{LB}, b_p^{UB})$ where $b_p^{LB} = ||_v -r^{LB}_{s,t}$, $b_p^{UB} = ||_v R^{UB}_{s,t}$. Note that $t$ is fixed here.
\par 
We also have the constraints that 
\begin{equation}
    0 \leq x \leq 1
\end{equation}
and the power balance across the zones:
\begin{equation}
    e^Tx = \sum_z \left(\sum_{b \in \mathcal{B}_z} Q_b^zx_b^z - \sum_{o \in \mathcal{O}_z} Q_o^z x_o^z\right) = 0
\end{equation}
where $e= 1^TE $. The primal problem then reads
\begin{align}
    \max_{x} &  \sum_{z \in \mathcal{Z}}\left( \sum_{b \in \mathcal{B}_z}Q_b^zP_b^zx_b^z  - \sum_{o \in \mathcal{O}_z}Q_o^zP_o^zx_o^z\right)= c^Tx\\
    s.t.& \hspace{5mm}  A_Ex \leq b_p  \hspace{4.5mm} (\gamma)\\
    &  \hspace{5mm} e^Tx = 0  \hspace{7mm} (y)\\
    & \hspace{5mm}  x \leq 1  \hspace{10.5mm} (p)\\
    &  \hspace{5mm} x \geq 0, \hspace{9mm} (w)
\end{align}
where $A_E =M_pE$. This gives us the maximal social welfare, but the price is not determined at this point.  To achieve this, the following dual problem is solved 
\begin{align}
    \min_{y,\gamma,p} & 0^Ty + r^T\gamma + 1^Tp\\
    s.t.  \hspace{5mm} & A_E^T\gamma + y e  + I(p-w) \geq c\\
    &  \hspace{5mm} \gamma \geq 0, p \geq 0, w \geq 0
\end{align}
and denote $\tilde{p} = p - w$. Once the dual variables $\gamma$ are known, the marginal transmission costs condition is imposed \cite{tarjei}:
\begin{equation}
    P_{z_i}- P_{z_j} = \sum_{k \in \mathcal{L}_{\cb}} \gamma_k (PTDF_{k,z_j} - PTDF_{k,z_i}), \hspace{5mm} j \neq i
\end{equation}
where $\mathcal{L}_{\cb}$ is the critical branches set. In matrix form, if we let $\Omega_Z \in \mathbb{R}^{|Z|(|Z|-1)/2 \times |Z|}$ be the matrix that takes all the pairwise differences between entries in zones we can write
\begin{equation}
    \Omega_z(\gamma^T M_p)^T = \Omega_z \Pi,
\end{equation}
where $\Pi$ is the vector of zonal prices. Note that this equation is actually under-determined since many constraints are linear combinations of the others. Thus, to ensure a unique zonal price, the price in each zone is in practice required to be close to the average between the lowest accepted bid and the highest accepted offer \cite{JAODocumentation}, such that we solve
\begin{align} 
    \min_{\Pi_z} & \sum_z \label{eq15} \left(\frac{P_O^z+P_B^z}{2}- \Pi_z\right)^2\\
    s.t. & \hspace{5mm}  \Omega_z(\gamma^TM_p)^T = \Omega_z \Pi, \label{eq16}
\end{align}
Where $P_O^z$ and $P_B^z$ are the highest offer and lowest bid accepted for zone $z$, respectively. 
\par
Once $\Pi$ is obtained, it is possible that some accepted bids have $x_b^z>0$ yet $\Pi_z>P_b^z$, and similarly, some offers may have $x_o^z>0$ yet $\Pi_z<P_o^z$ (these are known as paradoxically accepted bids and offers), and thus the set of bids and offers that remain valid needs to be re-evaluated. Further, note that in reality $\Pi$ can have negative entries, and this happens in practice since renewable plants bid at negative levels, since the price they receive is $\Pi_z +F$ where $F$ is a fixed positive fee. Thus, renewables may bid any price up to $-F$. If for example renewables can cover the full demand level, then the price is likely to be negative.
\par
\begin{definition}
A bid $b \in \mathcal{B}_z$ (or offer $o \in \mathcal{O}_z$) is said to be paradoxically accepted if when solving the social welfare problem, $x_b^z>0$ ($x_o^z>0$) but $P_b^z < \Pi_z$ ($P_o^z > \Pi_z$) where $\Pi_z$ is the unique zonal price determined by the pricing mechanism. 
\end{definition}
\par
To resolve the paradoxically accepted order issue, one could imagine solving the primal, dual and the pricing problem simultaneously. This could be achieved by using strong duality to turn the primal and dual objectives into a single equality constraint, and merge this problem with the pricing problem. However, we now have to enforce that paradoxical bids and offer do not exist which requires non-linear constraints of the form:
\begin{align}
    (P_b^z-\Pi_z)x_b^z \geq 0, \\
    (\Pi_z- P_o^z)x_o^z \geq 0,
\end{align}
resulting in bi-linear constraints that drastically increase problem difficulty. Using McCormick approximation for a sufficient condition generally gives infeasibility at least if the interval is not split in multiple sub-intervals, which sharply increases the computational cost.  Fortunately, we show that for the case of inflexible demand, paradoxically accepted orders do not exist.
\begin{theorem}\label{th322}
Let the social welfare maximisation problem, its dual and the pricing problem be defined as above. Then, given inelastic demand, the zonal prices $\Pi_z^* = \max_{o \in \mathcal{O}_z}\{P_o^z: x_o^z>0 \}$ are the optimal solution to the optimisation problem given by \eqref{eq15} and \eqref{eq16}, and $\Pi^*$ does not yield any paradoxically accepted orders for the primal problem.
\end{theorem}
The proof is provided in the Appendix for the interested reader, but this justifies further our modelling choice of inflexible demand, as a way to avoid the problem of paradoxically accepted orders, and ensures that the zonal prices are well defined.
%%       \citet{<label>} ==> Jones et al. [21]
%%       \citep{<label>} ==> [21]
%%
\section{A Social Welfare Maximisation market clearing model}\label{sec:SWMmodel}
We now propose a market clearing model that accurately represents the practice of market clearing procedure in what was the CWE market during the 2015-2021 period, currently extending to a 13 country market. To our knowledge, the presence and analysis of such optimisation models is very scarce in the literature. 
\par We begin by considering inflexible demand, which is a reasonable assumption in spot electricity markets. According to Theorem \ref{th322}, this allows us to avoid the problem of paradoxically accepted orders, while still retaining the main features of the market. Further, Theorem \ref{th322} gives us a simple way to compute the zonal prices that is consistent with the FBMC methodology, circumventing the need for solving a dual and a pricing problem. This greatly reduces computational cost. To further enhance computational tractability we neglect block order bids, and assume that producers have linear marginal cost 
\begin{equation}
    \frac{\partial C_i}{\partial x_i}(x_i) = c_ix_i+b_i
\end{equation}
where $C_i$ is the total production cost of producer $i$ when producing quantity $x_i$ and $c_i>0$ and $b_i>0$ are cost parameters. It is reasonable to assume that the order functions are in the same parametric class as the cost one. Thus, similarly to \citep{DRalph2007}, we consider linear marginal ask prices for the producers of the form
\begin{equation}
    \lambda_i(x_i) = m_i x_i + a_i,
\end{equation}
where $\lambda_i$ is the marginal price asked by producer $i$ when producing quantity $x_i$, while $m_i>0$ and $a_i>0$ are order constants. Note that for convenience, unlike in the introduction, $x_i$ represent actual quantities as opposed to fractions. The matrix $E$ then retains the same meaning but becomes populated by $\pm 1$ and $0$ entries only. We consider capacity constraints for each player, and forbid producers to be net short, that is we require $0 \leq x_i \leq Q_i$ $\forall i$, where $Q_i$ is the production capacity of producer $i$. Finally, each producer $i$ is assigned to the set of producers in that zone denoted as $\mathcal{P}_z$.
\par 
Network constraints have a linear form in the zonal production quantities $y = Ex$, and can be obtained as
\begin{equation}
    M_p y \leq b_p + M_p d^Z=:\overline{b}_p,
\end{equation}
where $d^Z$ is the vector of zonal demands. In practice these network constraints are obtained by following our work in \citep{PuiuHauser2021}, which for brevity this is not detailed further here. Supply and demand balance is enforced simply by $1^Ty = d$ (or $1^Tx = d$), where $d=1^Td^Z$ is the total demand.
\par 
As aforementioned, the market clearing mechanism for CWE Day Ahead Auction is based on the idea of \textit{Social Welfare Maximisation}. This defines the objective to maximise as the sum of total apparent benefits obtained by the consumer minus the total apparent costs of the producer. Unlike in the usual literature, we point out that the quantities used are \textit{apparent} because they neglect the possibility of un-truthful bidding and the effect of marginal pricing. Considering the form of the bids we assumed, and inflexible demand, the objective becomes:
\begin{equation}
\begin{aligned}
    F_{SWM} &= -\sum_i \int_0^{x_i} (m_i\chi_i+a_i)d\chi_i \\ &= -\sum_{i}\left[\frac{1}{2}m_i x_i^2+a_i x_i\right].
\end{aligned}
\end{equation}
It is worth noting that unlike for the case of Locational Marginal Prices with Optimal Power Flow, the SWM for inflexible demand does \textit{not} reduce to the total cost. This is because of zonal pricing, meaning that all producers in one receive the same price $\rho_z^*$, and not the individual highest asked and accepted one, $\lambda_i(x_i^*)$. The SWM optimisation problem can then be written as
\begin{equation}
    \centering
    \begin{aligned}
        \min_{x}&\sum_i\left(\frac{1}{2}m_i x_i^2+a_i x_i\right) = \frac{1}{2}x^TD_mx+a^Tx\\
        s.t.& \hspace{2mm} M_pEx \leq b_p\\
        &x_i \leq Q_i \hspace{1mm} \forall i\\
        &x_i \geq 0 \hspace{1mm} \forall i\\
        &1^Tx = d
    \end{aligned}
\tag{SWM}\label{SWM}
\end{equation}
Note that there are no price variables in either the objective or constraint set. As per the market solution in practice and according to Theorem \ref{th322}, the prices can be computed after solving the SWM problem and the largest price in each zone (with corresponding strictly positive production) is retained. Further, the optimal active player set is simply the set of players $\mathcal{P}^A:=\{i:x_i^*>0\}$, which can be trivially computed once the optimal solution of SWM is obtained. Note that our SWM model is a linearly constrained strictly convex quadratic program (CQP), which can be solved very effectively numerically. We solve this CQP corresponding to the SWM optimisation problem introduced here via the \textit{quadprog} package in \textit{Python 3}. 
\par Finally, note that while the considered structure of orders is linear in marginal asked prices, the optimisation model is agnostic to the cost structure itself, if strategic bidding is to be considered. We here make the usual assumption that the asks reflect the true costs, to enable us focusing on numerical solutions for the SWM and CM market clearing mechanism, as well as their comparison. We further find that even in this case by calibrating the SWM model, it can generally represent the real price data well. However, we consider strategic bidding in another work, where we also challenge the truthful bidding assumption.
\section{A Cost Minimisation model for market clearing}
We now introduce our proposed Cost Minimisation Market clearing. We arrive at the final optimisation model using the same assumptions and procedure as in Section \ref{sec:SWMmodel}, with the main difference being that the price variables now appear explicitly both in the objective and in the constraint set. The total electricity procurement cost to be minimised can be expressed as 
\begin{equation}\label{totalCost}
    C = \sum_z \rho_zy_z,
\end{equation}
where $\rho_z$ and $y_z = \sum_{i \in \mathcal{P}_z}x_i$ are the clearing price and total production quantities for zone $z$. By definition of our objective function, given fixed $y_z$, $\rho_z$ is the minimum price for which such clearing is ensured, thus
\begin{equation}\label{pricedef}
    \rho_z = \max_{j\in \mathcal{P}_z}\left\{\lambda_j(x_j):x_j>0 \right\}.
\end{equation}
The alert reader might observe that \eqref{pricedef} is an optimisation problem itself, and therefore the resulting cost minimisation market clearing model is in principle a bi-level optimisation problem. We later explain how this can be avoided both exactly and approximately, to only require solving a single level optimisation problem.
\subsection{The main model}
We use the assumption of inflexible demand, the same network constraints and producers' capacity constraints, and supply and demand balance requirement, to obtain the following Cost Minimisation bi-level optimisation problem with bilinear terms (bl-BLP):
\begin{equation}
\begin{aligned}
    \min_{x,y}& \sum_{z=1}^{N_z}\rho_{z}y_z,\\
    s.t. \hspace{3mm}& y_z = 1^Tx_z, \hspace{3mm} \forall z,\\
    &\rho_z = \max_{j \in \mathcal{J}_z} \{\lambda_j\left((x_z)_j \right): (x_z)_j>0 \},\\
    & \sum_{z =1}^{N_z} {y_z} =  \sum_{z =1}^{N_z} {d_z}^Z = d,\\
    &x_i \geq 0 \hspace{3mm} \forall i,\\
    &Q_i - x_i \geq 0 \hspace{3mm} \forall i,\\
    & M_p y \leq b_p,
\end{aligned}
\tag{CM-bl-BLP}\label{CM-bl-BLP}
\end{equation}
where $\lambda_i = m_ix_i +a_i$ is the bid price as a function of assigned production quantity $x_i$. With this formulation we have nested optimisation problems due to the $\max$ function in the equality constraint for $\rho_z$, which cannot be separated since $\rho_z$ appears in the objective. Due to the appearance of price variables in the objective and constraints, this optimisation problem is clearly much more challenging than its \eqref{SWM} counterpart, but comes with the added benefit of minimising the actual total procurement cost, therefore guaranteeing that for the same set of orders, this cost is lower when solving the CM model defined by \eqref{CM-bl-BLP}, than when solving the \eqref{SWM}. Further, as one would intuitively expect, the zonal prices are also generally decreased since these are part of the objective, and we confirm in Section \ref{sec:numres} that this is generally (albeit not always) the case.
\par 
Perhaps due to the much more challenging computational nature, to the best of our knowledge, the Cost Minimisation model received no attention in the literature directly concerned with the FBMC and CWE, and fairly little attention even for other markets, with the work of \citep{Alguacil2013} being perhaps the closest to our model. However, we show that strong analytical progress towards the solution can be made for our model, and efficient numerical computations are possible. In the next subsections we devise four efficient numerical algorithms, some of which have very strong convergence guarantees. This should be of interest to both policy makers, that have an interest in limiting producers' market power, as well as market participants that could use a CM model even under the current condition to obtain lower bound estimates of prices.
\subsection{The model under known active player set}
Part of the complexity when solving problem \eqref{CM-bl-BLP} stems from the fact that the active player set for each zone, $\mathcal{J}_z:=\{j \in \mathcal{P}_z:x_j>0\}$ is not known in advance. However, one could estimate the active player set by solving a simpler form of \eqref{CM-bl-BLP}, or even the  \eqref{SWM} problem. The latter would still guarantee that upon solving \eqref{CM-bl-BLP} with the known active player set, the total cost is lower than solving \eqref{SWM}. However, it also guarantees that the computational effort for (approximately) solving \eqref{CM-bl-BLP} is greater than the one for solving \eqref{SWM}, since extra computations are required after \eqref{SWM} is solved.
\par 
Let us now imagine that the active player set $\mathcal{J}_z$ for each zone $z$ is known, and observe how the CM model simplifies in this case. One can note that $\rho_z$ can be expressed as the minimum value greater than all marginal prices given by the accepted offers:
\begin{equation}
\begin{aligned}
    \rho_z = \min_{v_z} & \hspace{0.5mm}v_z\\
    s.t. \hspace{1mm} v_z\geq &\hspace{0.5mm}m_i x_i + a_i, \hspace{3mm} \forall i \in \mathcal{J}_z
\end{aligned}.\label{rhosys}
\end{equation}
Note that the active player set assumed is strict, in the sense that this formulation still requires that $x_i>0$, since $v_{z_i} \geq a_i$ is not required when $x_i = 0$, but this may occur if our estimate $\mathcal{\hat{J}}_z $ for $\mathcal{J}_z$ only guarantees that $\mathcal{J}_z 	\subset \mathcal{\hat{J}}_z$ (but not that $x_j^* =0 \implies x_j^* \notin \mathcal{\hat{J}}_z$). However, this formulation does allow for $v_{z_i}>m_ix_i+a_i$ that happens if $x_i = Q_i$, which may often be the case for at least a few producers. The market clearing bi-linear problem (mc-BLP) then becomes
\begin{equation*}
\begin{aligned}
    \min_{x,v,y}& \sum_{z=1}^{N_z}v_{z}y_z, \\
    s.t. \hspace{3mm}& y = E x,\\
    &v_z \geq m_i x_i + a_i \hspace{2mm} \forall i \in \mathcal{J}_z,\\
    & 1^Tx = d,\\
    & M_p E x \leq b_p, \\
    &x_i \geq 0, \hspace{3mm} \forall i,\\
    &Q_i - x_i \geq 0, \hspace{3mm} \forall i,
\end{aligned}    
\tag{mc-BLP}\label{mc-BLP}
\end{equation*}
since we have that $\min_{x,y}\sum_{z=1}^{N_z}\min_{v_z}\{v_{z}\}y_z =\min_{x,v,y}\sum_{z=1}^{N_z}v_{z}y_z  $
The zonal price is now denoted as $v_z$ instead of $\rho_z$, to distinguish the different price properties when compared to \eqref{CM-bl-BLP}. In this case, the matrix $E \in \mathbb{R}^{N_z \times N_p}$ maps the player quantities $x$ to the zonal quantities $y$ by summing all the produced quantities $x_i$ with $i \in \mathcal{J}_z$ in a zone $z$. The inactive players can thus be completely removed from the optimisation problem \eqref{mc-BLP}.
\par 
It is worth observing that if the zonal quantities $y$ were known, then the problem becomes an LP for $(x,v)$, since the only nonlinear (bi-linear) terms are the $v_zy_z$ ones in the objective. However, the problem is nonlinear in general and the computational cost for solving an optimisation problem with bilinear objective can be rather high.
\subsection{Solution Approaches}
In this section we devise four methods to solve \eqref{CM-bl-BLP} formulation of the Cost Minimisation spot model. Loosely speaking we present the methods in the order of increasingly stronger exploitation of problem structure, and for the first three, also based on convergence properties that come at the cost of increased complexity. The last method we propose (ib-CQP) is perhaps the most elegant, and makes the most use of analytical results. It perhaps offers the best trade-off between convergence speed and computational complexity, although strictly speaking it can only guarantee global convergence on sub-intervals generated by capacity constraints, instead of the whole space.
\par 
In our opinion, presenting all four methods is necessary due to different trade-offs between computational efficiency and convergence guarantees. Further, given modifications or generalisations of \eqref{CM-bl-BLP} the ideas of some methods might be more relevant than others, depending on the case. The application-oriented reader could skip to Sub-section \ref{sec:ib-CQP}, which we found to work best when applying our models to the real world case of CWE data, but as a word of caution we mention that this does not have to always be the case, although the ib-CQP method has desirable properties.
\subsubsection{Iteratively enhanced quasi LP (ieq-LP) approach}
This approach observes that if $y$ was approximated just in the objective, then \ref{mc-BLP} becomes an LP. This observation is important because the zonal production vector $y$ is much lower dimensional, and fractional changes are expected to be lower than $x$ values, when the objective is approximated. Further, the constraint $y=Ex$ becomes redundant and can be removed. Thus, given a guess value for the zonal quantities, $\hat{y}$ we solve the following LP:
\begin{equation*}
\begin{aligned}
    \min_{x,v}& \sum_{z=1}^{N_z}v_{z}\hat{y}_z, \\
    s.t. \hspace{3mm} &v_z \geq m_i x_i + a_i, \hspace{2mm} \forall i \in \mathcal{J}_z,\\
    & 1^Tx = d,\\
    & M_p E x \leq b_p, \\
    &x_i \geq 0, \hspace{3mm} \forall i,\\
    &Q_i - x_i \geq 0, \hspace{3mm} \forall i.
\end{aligned}    
\tag{mc-qLP}\label{mc-qLP}
\end{equation*}
once \ref{mc-qLP} is solved, we update $\hat{y}=Ex^*$, and repeat the process until converged. Note that it is key to remove the constraint $\hat{y}=Ex$, to allow for updates in $\hat{y}$, for else this is constant and an iterative approach cannot improve the solution. This approach is denoted as iterative quasi-LP (iq-LP). 
\par 
Unfortunately, this approach is not guaranteed to converge and it can be observed that the actual objective function ${v^*}^TEx^*$ can increase from one iteration to another. To resolve this problem, instead of retaining the solution given by the last iteration, the solution with the minimum actual objective function, ${v^*}^TEx^*$, is always stored. To resolve the convergence issue a maximum number of iterations is imposed. This is described in Heuristic \ref{alg:ieq-LPh}.
\begin{heuristic}[h!]
\caption{Iterative enhanced quasi LP (ieq-LP)}\label{alg:ieq-LPh}
\begin{algorithmic}
\Require input $(m_i,a_i)$ for all $i \in \cup_z\mathcal{J}_z$, maximum number of iterations $N^q_{max}$, and tolerance $\delta$
\State Set $y^0 \gets d$ or the Chebychev centre of the $(M_p,b_p)$ polytope.
\State Initialise: $x,v,y \gets \mathrm{None}$ and $f_{best} \gets \infty$
\For{$i \in \{1,...,N_{max}^q\}$}
 \State obtain $x^i, v^i$ by solving \ref{mc-qLP}($y^{i-1})$
 \State obtain $y^i \gets Ex^i$ (or $y^i \gets \alpha_y y^{i-1} + (1-\alpha_y)Ex^i$)
  \If{${v^i}^TE x^i< f_{best}$}
 \State $f_{best} \gets {v^i}^TE x^i$
 \State $x \gets x^i$, $v \gets v^i$, $y \gets y^i$
\EndIf
 \If{$ \|x^i -x^{i-1} \|^2_2/ \|x^i \| < \delta$}
 \State return $x, v, y, f_{best}$
\EndIf
\EndFor
\State return $x$, $v$, $y$, $f_{best}$
\end{algorithmic}
\end{heuristic}
The aim of Heuristic \ref{alg:ieq-LPh} is therefore not to obtain a global or even local minimizer, but rather to quickly obtain a feasible solution that has an objective function \textit{"as small as possible"}. We also considered the case of updating $y^i$ as $y^i \gets \alpha_y y^{i-1} + (1-\alpha_y)Ex^i$, with $0<\alpha_y<1$ which is an exponentially moving average approach, and we denote the corresponding algorithm as (ieq-LP-ma). We found this variant to work better in practice since it avoids cycling between different points by changing $y^i$ \textit{"sufficiently slow"}.
\subsubsection{Iterative ellipsoids for Quadratic programs (ie-QP) approach}
If we define $z^T = (x^T, v^T)$, and assuming that the set of active players $\mathcal{J}_z$ for all $z$ is known, \eqref{CM-bl-BLP} can be recast as 
\begin{equation*}
\begin{aligned}
    \min_{z}& z^T G z \\
    s.t. \hspace{3mm}&
    A_I z \leq b_I \\
    & \overline{e}_E^T z = d,
\end{aligned}    
\tag{mc-QP}\label{mc-QP}
\end{equation*}
where $d$ is the total demand, $\overline{e}_E^T = (e_x^T, 0^T)$, with $e_x \in \mathbb{R}^{N_p}$ the vector of all ones, and
\begin{equation*}
    G = \begin{bmatrix}
    0 & \frac{1}{2}E^T\\
    \frac{1}{2}E & 0
    \end{bmatrix}; \hspace{2mm}
    A_I = \begin{bmatrix} D_m & -E^T \\
    -I & 0 \\
    I & 0 \\
    M_pE & 0
    \end{bmatrix}; \hspace{2mm} b_I = \begin{bmatrix} -a\\ 0 \\ Q \\ b_p 
    \end{bmatrix},
\end{equation*}
where $D_m = \textrm{diag}(m)$. One could also introduce regularisation by replacing $G$ by some $\hat{G}  = G + \textrm{diag}(\eta)$ for some vector $\eta>0$. The key idea is to recognise that the set of inequailties $ A_I z \leq b_I$ defining the space $\mathcal{Z}_p : =\{z:A_I z \leq b_I \} $ can be replaced by a stricter one of ellipsoid form, $\mathcal{Z}_e:=\{z: \|A_ez-p_e\|_2^2 \leq r_e^2 \} \subseteq \mathcal{Z}_p$, for some $A_e$, $p_e$ and $r_e$. Further let us assume that $\{z:M_ez=b_e\} \subseteq \{z:\overline{e}_E^T z = d\}$ for some matrix $M_e$ and vector $b_e$. As a consequence, the optimal point $z^{*,e}$ of
\begin{equation*}
\begin{aligned}
    \min_{z}& z^T \hat{G} z \\
    s.t. \hspace{3mm}&
   \|A_ez-p_e\|_2^2 \leq r_e^2 \\
    & M_e z = b_e,
\end{aligned}    
\tag{em-TR}\label{em-TR}
\end{equation*}
must be feasible for \eqref{mc-QP}. Further, if $\|A_ez^{*,e}-p_e\|_2^2 < r_e^2$, then $z^{*,e}$ is also a local minimum for \eqref{mc-QP}. On the other hand if $\|A_ez^{*,e}-p_e\|_2^2=r_e^2$, one could define another ellipsoid satisfying $\mathcal{Z}_{e'} \subseteq \mathcal{Z}$ and centred at $z^{*,e}$ to allow for future progress. This procedure can then be repeated until a strictly interior point is found to be a minimum, or until ellipse size becomes very small. There are however two issues to address next: (i) how to solve \eqref{em-TR} and how to obtain $\mathcal{Z}_e:=\{z: \|A_ez-p_e\|_2^2 \leq r_e^2 \}\subseteq \mathcal{Z}_p$ constraints. We first start by addressing solving \eqref{em-TR}, which can be done by
observing that this minimisation problem is a trust-region like problem \citep{NoceWrig06}, with ellipsoid instead of spherical trust region, and an extra equality constraint. However, these extra features do not change the procedure of the solution. The first step is writing the KKT system as 
\begin{align*}
    &2\hat{G}z  + M_e^T\gamma - \lambda(2A_s^TA_s z - 2A_s^Tp_s) =0,\\
    &M_e z= b_e,\\
    &\|A_s z-p_s \|_2^2 \leq r_s^2,\\
    &\lambda(\|A_s z-p_s \|_2^2  - r_s^2) =0, \hspace{2mm} \lambda \leq 0.
\end{align*}
The equality constraint $\lambda(\|A_s z-p_s \|_2^2  - r_s^2) =0$ tell us that there are only two possibilities: either $\lambda =0$, or $\|A_s z-p_s \|_2^2  - r_s^2=0$ (or both).  If $\lambda =0$ then we have that
\begin{equation}\label{unconstrained_y_ellipse}
    \begin{pmatrix}
z^*\\ \gamma^*
\end{pmatrix} = \begin{pmatrix}
2\hat{G} & M_e^T\\
M_e& \mathbf{0}
\end{pmatrix}^{-1}\begin{pmatrix}
0\\ b_e
\end{pmatrix}.
\end{equation}
Now if $ \|A_ez^{*,u}-p_e\|_2^2 \leq r_e^2$ then a solution has been found. However, \eqref{em-TR} will always yield unbounded solution if $\eta=0$, and then solving \eqref{unconstrained_y_ellipse} can be skipped, and the next stage can be directly applied. 
\newline
If $ \|A_ez^{*,u}-p_e\|_2^2> r_e^2$ then we have not obtained a feasible solution $z^{*,u}$, but we can conclude that the solution has to be on the boundary, i.e. $ \|A_ez^{*,u}-p_e\|_2^2 = r_e^2$. In this case we can write
\begin{equation}\label{constrained_y_ellipse}
    \begin{pmatrix}
z\\ \gamma
\end{pmatrix} = \begin{pmatrix}
2\hat{G} - 2\lambda A_s^TA_s & M_e^T\\
M_e& \mathbf{0}
\end{pmatrix}^{-1}\begin{pmatrix}
-2A_s^Tp\lambda\\ b_e
\end{pmatrix},
\end{equation}
and let us denote  
\begin{equation*}
     \hat{D}(\lambda) := \begin{pmatrix}
2\hat{G} - 2\lambda A_s^TA_s & M_e^T\\
M_e& 0 \end{pmatrix}; \hspace{3mm} u(\lambda) := \begin{pmatrix}
-2A_s^Tp_s\lambda\\ b_e
\end{pmatrix}.
\end{equation*}
Since we know that $\|A_sz-p_s \|_2^2= r^2$ and thus we have that $f(\lambda) = \frac{1}{||A_0\hat{D}(\lambda)^{-1} u(\lambda)-p||_2^2} - \frac{1}{r^2} = 0$, which is a one dimensional problem that can be solved with Newton-Rapson.
\par 
All we have left to do is to obtain $\mathcal{Z}_e$ sets. These ellipses can be obtained by writing the log barrier function based on the linear constraints
\begin{equation}
    f_b(z) = -\sum_k \ln\left( (b_I)_k - (M_I)_{k,:}z\right),
\end{equation}
which is only defined inside the polytope $\mathcal{Z}_p$. The hessian of this function can be obtained as
\begin{equation}
    H_b(z) = \nabla^2f_b = -M_I^T \textrm{diag}(g_b(z))^2 M_I,
\end{equation}
where $g_b(z) := \nabla f_b(z)$ with 
\begin{equation}
    (g_b)_k(z) = \frac{1}{(b_I)_k - (M_I)_{k,:}z}.
\end{equation}
As per \cite{JamesRenegar}, if $y_c$ is an interior point of $\mathcal{Z}_p$, the ellipse
\begin{equation}
    EL(z_c, M_I, b_I):=\{ z: (z-z_c)^TH_b(z_c)(z-z_c) \leq 1\} 
\end{equation} 
is guaranteed to lie inside the polytope $\mathcal{Z}_p$. One can now simply use this ellipsoid constraint in \eqref{em-TR}, or can reduce it to the standard form by performing the Cholesky decomposition $H_b = LL^T = A_s^TA_s$, and so $A_e = L^T$, $p_e = A_ez_c$, $r_e=1$, and the ellipse can then be re-written as
\begin{equation}
    EL(z_c, M_p, b_p):=\{ z: \|A_s z - p_s\|_2 \leq r_e \}.
\end{equation} 
Let us denote this procedure of obtaining $A_s$, $p_s$ and $r$ based on $y_c$, $M_p$, $b_p$ as $\textrm{EL}(y_c, M_p, b_p)$. Note that there exist points in $\mathcal{Z}_p$ that are not in $EL(z_c, M_p, b_p)$, and thus by replacing $\mathcal{Z}_p$ with $EL(z_c, M_p, b_p)$, the obtained solution may not be a local minimum on $\mathcal{Z}_p$. This approximation becomes increasingly worse as the dimensionality of $z$ increases. 

\begin{algorithm}[h!]
\caption{ie-QP with reconditioning (ie-QP-wr)}\label{alg:ie-QP-wr}
\begin{algorithmic}
\Require input $(m_i,a_i, Q_i)$ for all $i \in \cup_z\mathcal{J}_z$, maximum number of iterations $N_{max}$, total demand $d$, convergence tolerance $\delta$, and boundary tolerance $\delta_B$.
\State Get the Chebychev centre $z_c$ of the $(A_I,b_I)$ polytope.
\State initialise $M_e \gets (\mathbf{1}_{N_p}^T, \mathbf{0}_{N_z}^T)$, $b_e \gets d$
\For{$i \in \{1,...,N_{max}\}$}
 \State obtain $A_e,p_e,r_e \gets \textrm{EL}(z_c,A_I,b_I)$ 
 \State obtain $z$ by solving the \eqref{em-TR} problem, using ellipsoid $(A_e,p_e,r_e)$
 \State compute $\Delta  z = \|z_i^*-z_c\|_2$ 
  \If{$ \Delta z< \delta $}
 \State return $z_i^*$, $\Delta z$
\EndIf
 \State $z_c \gets z_i^*$
 \State Let $\mathcal{L} := \overline{1,\textrm{dim}(b_I)}$ $\mathcal{E}: = \{l: (b_I)_l-(A_I)_{l,:}z_i^*< \delta_B, l \in \mathcal{L}\}$
 \State $A_I \gets A_I[\mathcal{L}\setminus\mathcal{E},:]$, $b_I \gets b_I[\mathcal{L}\setminus\mathcal{E}]$
 \State $M_e \gets ||_v(M_e,A_I[\mathcal{E},:])$ , $b_e \gets ||_v(b_e,b_I[\mathcal{E}])$
 \If{$\textrm{rank}(M_e)=\textrm{dim}(z)$}
 \State $z \gets (M_e^TM_e)^{-1}M_e^Tb_e$, $\Delta  z \gets \|z_i^*-z_c\|_2$
 \State return $z_i^*$, $\Delta z$
 \EndIf
\EndFor
\State return $z^*_i$, $\Delta z$
\end{algorithmic}
\end{algorithm}

\par 
We can control the approximation error, or indeed reducing it to zero, by generating a sequence of ellipses as following: (i) given an interior point $z_c$, compute the ellipse $\textrm{EL}(z_c, M_p, b_p)$, (ii) solve the corresponding \ref{em-TR} on this ellipse to obtain a solution $z^{*,e}$, and (iii) replace $z_c$ by $z^{*,e}$ and go to (i), or stop if a specific convergence criterion is met. 
\par 
However, in practice we observe that the ellipsoid constraints of the form $\|A_e z - p_e\|_2 \leq r_e$ can have an increasingly ill conditioned $A_e$ as the solution iterate $z^*_l$ approaches the boundaries of $A_I z \leq b_I$. This results in very little progress per outer iteration and longer solution times for an individual Trust-Region problem. To resolve this problem, when inequality constraints become \textit{"close to equalities"} we transform these constraints to equality constraints, and search for solutions in the resulting subspace. Finally, if the equality system $M_ez=b_e$ becomes full rank at any iteration, there is no need to continue, as an unique solution exists, and this can be computed as $z^* = (M_e^TM_e)^{-1}M_e^Tb_e$. The final procedure is described in Algorithm \ref{alg:ie-QP-wr}.
\subsubsection{A Branch and Bound Tree (BBTree) approach}
The intuition behind using a Branch and Bound approach can be understood by looking at \eqref{mc-BLP} and observing that if $y$ was known, the problem becomes an LP which is easy to solve. Thus, one might consider trying \textit{all the possible values} for $y$, solve the corresponding LP with fixed $y$, and retain the solution with the minimum objective value. However, since $\mathcal{P}(y)=\{y:M_py \leq b_p\}$ is a continuous space, the number of values of $y$ to try is infinite. When discretising even just for $N_{dc}=10$ points per zone and for $N_z =5$, and the number of number of problems to be solved is $N_{dc}^{N_z}=10^5=100,000$ which is prohibitively large. 
\par 
Therefore, we propose a Branch and Bound Tree (BBTree) based approach that leverages more information about the problem to \textit{try to group similar y subdomains together}, such that early elimination of sub-optimal $y$ values is possible. For the BBTree approach the following are required:
\begin{enumerate}
     \item a method to compute an upper bound (UB) of the optimal solution on the feasible (sub-)space given,
    \item a method to compute a lower bound (LB) on the given feasible (sub-)space,
    \item a method to branch (split) the feasible (sub-)space considered into at least two regions, ideally in such a way to facilitate rapid elimination of sub-optimal (sub-)spaces,
    \item and a rule to choose the next tree node (defined by its feasible subspace), which is then to be processed.
\end{enumerate}
\par
\textbf{Obtaining an Upper Bound (UB)} on a feasible subspace given by $(M_p^j,b_p^j)$ at node $j$ is achieved by using Heuristic \ref{alg:ieq-LPh} (ieq-LP). Since the resulting solution is feasible, then the optimal objective can only be less or equal. This approach is used because it is fast, and it has been observed that often times it obtains a local or even global minimum. While this is not a guarantee, it is a useful property, since Heuristic \ref{alg:ieq-LPh} (ieq-LP) is called many times inside BBTree, once for each node, and therefore achieving a tight UB often, helping with pruning. Using Algorithm \ref{alg:ie-QP-wr} (ie-QP) to obtain an UB was also considered, but the computational time is significantly increased, and the increased UB tightness does not compensate for this, generally yielding a much slower BBTree when using ie-QP when compared to ieq-LP. The upper bound for node $j$ in the tree is thus updated as $f_{UB}(j) \gets f_{best}(\textrm{ieq-LP}(j))$, and the corresponding quantities and prices returned by ieq-LP$(j)$ are denoted as $x_{sol}(j)$, $v_{sol}(j)$, $y_{sol}(j)$.
\par
\textbf{Obtaining a Lower Bound (LB)} that is fast to compute and tight is rather challenging since it requires solving a (sufficiently) simplified problem that guarantees a (tight) lower bound. Multiple approaches were tried, but a variation of McCormick envelopes \cite{Mccormick_lecture_ref} seemed to perform best. To obtain the McCormick envelopes, we first need to compute $v_z^{min}$, $v_z^{max}$, $y_z^{min}$, $y_z^{max}$, for each zone $z$, and this can be achieved by solving the LP obtained by replacing the objective in \eqref{mc-BLP} with $v_z$, $-v_z$, $y_z$ and $-y_z$. The resulting problems are indeed LPs since both the objective and constraints are linear and we denote them by {\LPvm,} \LPvM, \LPym, \LPyM { respectively}, and let us denote by \LPVY{ the} process of computing all four quantities. Note that the problems for $y^{min}$ and $y^{max}$ can be further simplified by using that $Ex = y$ and removing all the other constraints for $x$, as well as removing $v$. The McCormick envelopes can now be obtained by using that given
\begin{align}
    v_z^{min}& \leq v_z \leq v_z^{max}\\
    y_z^{min}& \leq y_z \leq y_z^{max},
\end{align}
we have that
\begin{align}
    v_zy_z \geq &v_z^{min}y_z +y_z^{min}v_z - v_z^{min}y_z^{min}=: w_z^{LB,1}\\
    v_zy_z \geq &v_z^{max}y_z +y_z^{max}v_z - v_z^{max}y_z^{max} =: w_z^{LB,2},
\end{align}
and so by summing over all zones we get that
\begin{equation}
    \sum_zv_zy_z \geq w_z^{LB,1}; \hspace{3mm}\sum_zv_zy_z \geq w_z^{LB,2}, 
\end{equation}
and by taking the maximum 
\begin{equation}
    \sum_zv_zy_z \geq \max  \left\{\sum_z w_z^{LB,1},\sum_z w_z^{LB,2} \right\}.
\end{equation}
A tighter bound could be obtained as 
\begin{equation}
    \sum_zv_zy_z \geq \max_{i_z \in \{1,2\}} \sum_z w_z^{LB,i_z},
\end{equation}
but then the resulting problem is a Mixed Integer Linear Program (MILP) instead of LP, which is much harder to solve. Note that $v_z^{min}$, $v_z^{max}$, $y_z^{min}$ and $y_z^{max}$ are a function of $M_p^j$ and $b_p^j$ and thus are updated for each node $j$. Given the bounds on $y$ and $v$, one can now compute the lower bounds on the objective over the domain $(M_p^j, b_p^j)$ as
\begin{equation*}
\begin{aligned}
    \min_{x,v,y, w^{LB,l}}& \sum_{z=1}^{N_z}w_z^{LB,l}\\
    s.t. \hspace{3mm}& y = E x,\\
    &v_z \geq m_i x_i + a_i \hspace{2mm} \forall i \in \mathcal{J}_z,\\
    & 1^Tx = d,\\
    & M_p E x \leq b_p, \\
    &x_i \geq 0 \hspace{3mm} \forall i,\\
    &Q_i - x_i \geq 0 \hspace{3mm} \forall i,
\end{aligned}    
\tag{Env-LP$(l)$}\label{Envelope-BLP}
\end{equation*}
for $l \in \{1,2\}$. The lower bound for node $j$ is then updated as $f_LB(j) = \max  \left\{\sum_z w_z^{LB,1},\sum_z w_z^{LB,2} \right\}$.
\par 
\textbf{Branching at a Node} is a key component to facilitate fast pruning and therefore convergence speed. Branching is achieved in three steps, as explained below:
\begin{enumerate}
    \item Compute the Chebyshev centre $y_c(j)$ of node $j$, and the corresponding radius $r_c(j)$. This point is chosen such that the size of the two resulting branches are balanced.
    \item Determine a cutting plane, which is chosen by finding a \textit{"global"} descent direction $s_D(j)$ and cutting at $y_c(j)$ via a hyperplane perpendicular to it. The intuition behind choosing such a direction is that if the objective function is approximately linear, then the branch including all points containing negative directions along $s_D(j)$ should be pruned fast, as its LB is expected to be higher than the UB of the other branch.
    \item Perform the cut and form the new branches.
\end{enumerate}
The mathematics required to perform these steps is now explained. Finding the Chebyshev centre amounts the simple LP
\begin{equation*}
\begin{aligned}
    \max_{y_c,r_c \geq 0}& \hspace{1mm}r_c \\
    s.t. &\hspace{1mm} M_p y_c + r_c w_m \leq b_p
\end{aligned}.
\tag{Cheb-C}\label{Cheb-C}
\end{equation*}
The \textit{"global descent"} direction is obtained as 
\begin{equation}
    s_D = \frac{1}{\|y_{sol}(j) - y_c(j) \|}\left(y_{sol}(j) - y_c(j)\right).
\end{equation}
Plane cutting is then performed by observing that the domains to be split can be expressed as
\begin{equation}
\begin{aligned}
    \mathcal{P}^j_1 := \{ y: (y-y_c(j))^Ts_D \geq 0\};\\ \mathcal{P}^j_2 := \{ y: (y-y_c(j))^Ts_D \leq 0\},
\end{aligned}
\end{equation}
where the subscripts on $\mathcal{P}^j$ refer to child $1$ and $2$ of node $j$. This results in requiring to add the following constraint line to each of the two branches:
\begin{equation}\label{computechildren}
    M_p^j(l) \begin{bmatrix} M_p^j\\
    (-1)^ls_D(j)^T\end{bmatrix}; \hspace{2mm} b_p^j(l) = \begin{bmatrix} b_p^j\\
    (-1)^ls_D(j)^Ty_c(j)\end{bmatrix}, 
\end{equation}
for $l=\{1,2\}$.
\par 
\textbf{Choosing the next node to process} has the following objectives: (i) not to neglect any leaf for a very long number of iterations, relative to the number of active leaves generated before it, (ii) to reduce the global UB, (iii) to increase the global LB, to (iv) explore new leaves to avoid getting stuck in a bad pattern. To this end, we perform node choice as following:
\begin{itemize}
    \item choose the oldest active leaf, $60\%$ of the time. The oldest active leaf is expected to be amongst the largest based on feasible domains, and thus it might include the global minimum with higher likelihood.
    \item With $10\%$ probability, choose the node that contains the best available solution (incumbent) to be processed. If ieq-LP achieves solutions at least as good as the parent for the children, the incumbent must be a leaf. Non-leaf incumbent was almost never observed, but our implementation safeguards against branching already branched nodes.
    \item With $20\%$ probability, choose the node with the lowest global LB. We observe in practice that the UB is significantly better than the LB, and thus more time is spent to increase the LB.
    \item With $10\%$ probability, choose uniformly at random an active leaf.
\end{itemize}
We call this rule the Randomised Node Selection rule (RNS). The pseudocode for the BBTree is presented in Algorithm \ref{alg:BBTree}. 
\begin{algorithm}[h!]
\caption{Compute Node Quantities (CNQ)}\label{alg:CNQ}
\begin{algorithmic}
\Require input $(m_i,a_i)$ for all $i \in \cup_z\mathcal{J}_z$, $(M_p^j,b_p^j)$, and domain size tolerance $r_{\delta}$
\State Compute the Chebychev centre $y_c(j)$ and the radius $r_c(j)$ of the $(M_p,b_p)$ polytope.
\State Compute $x_{UB}$, $v_{UB}$, $y_{UB}$ and $f_{UB}$ as per ieq-LP, given $(M_p^j, b_p^j)$ as polytope constraints. The solution variables are also updated to this variables.
  \If{$r_c<r_{\delta}$}
 \State Set LB values to the UB values (since the distance between $y_{sol}$ and $y_{opt}$ must be very small)
 \Else
 \State for all $z$: $(v_z^{min}, v_z^{max}, y_z^{min},y_z^{max}) \gets$\LPVY
 \State Compute $x_{LB}$, $v_{LB}$, $y_{LB}$, $f_{LB}$ via \eqref{Envelope-BLP}, with $l=\{1,2\}$
\EndIf
\State $s_D(j) \gets  \frac{1}{\|y_{sol}(j) - y_c(j) \|}\left(y_{sol}(j) - y_c(j)\right)$
\State get branching domains $\{M_p^j(l), b_p^j(l)\}_{l\in\{1,2\}}$ as per \eqref{computechildren}
\State Store all computed quantities at node level
\end{algorithmic}
\end{algorithm}

\begin{algorithm}[h!]
\caption{Branch and Bound Tree (BBTree)}\label{alg:BBTree}
\begin{algorithmic}
\Require input $(m_i,a_i,Q_i)$ for all $i \in \cup_z\mathcal{J}_z$, demand vector $d$, $E$, $(M_p,b_p)$,  tolerance $r_{\delta}$, optimality gap $\delta_o$
\State Initialise root $\mathcal{R}_n$ of tree $\mathcal{T}$, on whole domain by calling CNQ($M_p,b_p$)
\State Set the active set of leaves $\mathcal{L}_A =\{ \mathcal{R}_n\}$ and $\delta_T = \infty$
\While {$\delta_T> \delta_o$}
\State choose node pointer $p$ to branch at (via RNS rule),
\State use CNQ to process both children nodes generated by pointer $p$
\State $\mathcal{L}_A \gets \mathcal{L}_A \cup \{c_p^1\} \cup \{c_p^2\} \rsetminus \{p\} $, where $c_p^l$ are the children of $p$.
\State update global upper bound $f_{UB}^G \gets \min\{f_{UB}(j):j \in \mathcal{L}_A\}$, and retain best node $s_n$
\State update global lower bound $f_{UB}^G \gets \min\{f_{LB}(j):j \in \mathcal{L}_A\}$ and retain lowest lower bound node $l_n$
\State for all active leaves $j \in\mathcal{L}_A$ remove node $j$ from active leaves if $f_{LB}(j)>f_{UB}^G$.
\State $\delta_T \gets 2(f_{UB}^G - f_{LB}^G)/(f_{UB}^G + f_{LB}^G)$
\EndWhile
\State fetch $x_{best}$, $y_{best}$, $v_{best}$, $f_{best}$ based on best current node $s_n$,
\State return $x_{best}$, $y_{best}$, $v_{best}$, $f_{best}$, $\delta_T$
\end{algorithmic}
\end{algorithm}
\subsubsection{Iteratively bounded CQPs (ib-CQP) approach}\label{sec:ib-CQP}
The intuition behind this method is that for each zone, the price function $v_z(y_z^*)$ is piecewise linear at optimality. This can be understood as following: if there are no capacity constraints, it is fairly easy to show that $v_z^* = \alpha_z y_z+\beta_z$, for 
\begin{equation}\label{alphabeta}
    \alpha_z = \left(\sum_{j \in \mathcal{J}_z} \frac{1}{m_j} \right)^{-1}; \hspace{2mm} \beta_z = \alpha_z \left(\sum_{j \in \mathcal{J}_z} \frac{a_j}{m_j} \right) .
\end{equation}
Thus, the same must be true if solving over regions of $x \in \mathcal{X}$ (that map to regions over $y \in \mathcal{Y}$), where no capacity constraints change state - that is, player sets $\mathcal{P}_z^I:=\{i \in\mathcal{Z}:x_i=0\}$, $\mathcal{P}_z^M:=\{i \in\mathcal{Z}:0<x_i<Q_i\}$ and  $\mathcal{P}_z^F:=\{i \in\mathcal{Z}:x_i=Q_i\}$ remain unchanged over these sub-domains of $\mathcal{X}$. For such a sub-domain, players in $\mathcal{P}_z^I$ can be simply disregarded, while players in $\mathcal{P}_z^F$ can be eliminated by reducing the production levels as $\hat{y}_z= y_z - \sum_{i \in \mathcal{P}_z^F}Q_i$. When these three player sets change, a change may occur in the price function, but this still remains linear between change points. However, by considering $v(y)$, it is not trivial to devise a method to directly find the three $\mathcal{P}_z^.$ sets for each $y$ in a computationally efficient way. For this reason we propose an alternate view that enables making analytical progress: instead of trying to compute $v(y)$ for fixed y at optimality, let us fix $v$, and compute the corresponding optimal $x$ and $y$. This might seem naive or counter-intuitive at first, as $v$ is perhaps the variable that has the greatest impact on the objective.
\begin{figure}[ht]
    \centering
    \includegraphics[width=7.9cm]{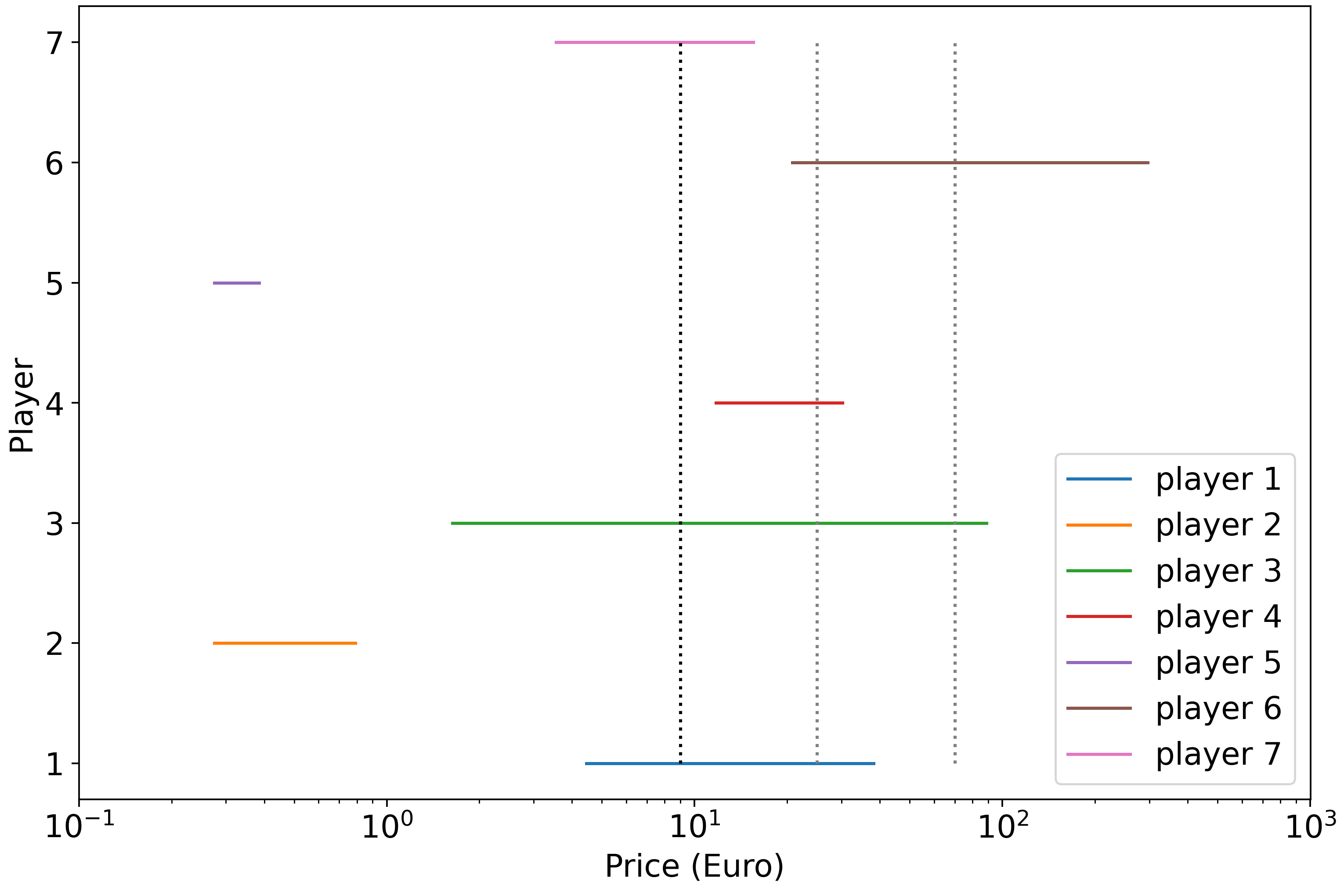}
    \caption{Obtaining the optimal price function: visualisation of player sets identification.}
    \label{fig:swipestuff}
\end{figure}
\par
However, this way of thinking allows us to make the following observation: for each zone, if the corresponding $v_z^*$ is the optimal value, then it has to be that (for $\forall i \in \mathcal{Z}$): (i) a player has $a_i > v_z^*$ if and only if it belongs to the inactive set, $i \in P_z^I$, (ii) a player does not sell at full capacity ($i \in \mathcal{P}_z^M$) if and only if $v_z = \lambda_i$, and (iii) a player sells at full capacity ($i \in \mathcal{P}_z^{F}$) if and only if $v_z\geq\lambda_i$. Note that the exact location of placing the equality signs is somewhat irrelevant. By computing the price that a player asks for at maximum capacity, $\lambda_i^{max}=m_iQ_i+a_i$ we can re-state the result in the following simplified form. At optimality we must have that (for $\forall i \in \mathcal{Z}$): (i) $v_z<a_i$ if and only if $x_i=0$, (ii) $a_i\leq v_z <m_iQ_i+a_i$ if and only if $0<x_i<Q_i$, and (iii) $v_z \geq m_iQ_i+a_i$ if and only if $x_i = Q_i$. This makes identifying the three player sets trivially for each $v_z$. The value of $v_z$ can then be swiped across $[0, \max_{i \in \mathcal{Z}} \lambda_i^{max}]$ to obtain all discontinuity points, as shown in Figure \ref{fig:swipestuff}. In fact, one can simply get all price points $\mathcal{V}_z:=\{a_i:i \in \mathcal{Z}\} \cup \{m_iQ_i+a_i:i \in \mathcal{Z}\} \cup \{0\}$, sort them and then compute the player set for each interval only when required. Now, for each $\nu_z \in \mathcal{V}_z^D$, one can obtain the corresponding $y_z$ value as:
\begin{equation}
    y_z(\nu_z) = (\nu_z-\beta_z(\nu_z))/\alpha_z(\nu_z) + \sum_{i \in \mathcal{P}_z^F(\nu_z)}Q_i,
\end{equation}
where $\mathcal{P}_z^F(\nu_z):=\{i \in \mathcal{Z}:\nu_z>m_iQ_i+a_i \}$ is the player set in zone z that has the maximum offered price below the current price $\nu_z$, and $\alpha_z(\nu_z))$ and $\beta_z(\nu_z))$ are computed as per \eqref{alphabeta}, by assuming the player set to be $\mathcal{J}_z \gets \mathcal{P}_z^M(\nu_z):=\{i \in \mathcal{Z}:a_i<\nu_z<m_iQ_i+a_i\}$. Let us denote the point set of all $y_z(\nu_z)$ corresponding to $\mathcal{V}_z^D$, $\mathcal{Y}^D_z$. Let us further denote the total production offered by players in zone $z$ at full capacity as $Q_z^s(\mathcal{Y}^D_z)=  \sum_{i \in \mathcal{P}_z^F(\nu_z)}Q_i$. Note that there is a one-to-one mapping between the two sets. Thus, for each $y_z$, we can find $y_z^L$ and $y_z^U$ such that $y_z^L\leq y_z <y_z^U$, and let us denote this interval $\mathcal{Y}^{I}_z(y_z)$. Further we can now obtain the $\alpha_z(\mathcal{Y}^I_z(y_z))$ and $\beta_z(\mathcal{Y}^I_z(y_z))$ corresponding to the $\mathcal{Y}^I_z(y_z)$ interval. Note that $\alpha_z(\mathcal{Y}^I_z(y_z))$ and $\beta_z(\mathcal{Y}^I(y_z))$ are fixed on each interval. This gives us the outstanding property that the cost minimisation objective $\sum_zv_zy_z$ is in fact piece-wise quadratic. Given any $\mathcal{Y}^I= \cup_z\mathcal{Y}^I_z$, the objective function can be exactly expressed for any $y \in \mathcal{Y}^I$ as
\begin{equation}\label{remarkableeq}
    \sum_{z=1}^{N_z}\alpha_z(\mathcal{Y}^I_z)y_z^2 + \left[\beta_z(\mathcal{Y}^I_z)-\alpha_z(\mathcal{Y}^I_z) Q_z^s(\mathcal{Y}^I_z)\right]y_z
\end{equation}
when $x_z$ are optimally chosen s.t. $x_z$ satisfies $1^Tx_z=y_z$ and the capacity constraints. This is a rather remarkable result that allows us to solve strictly convex quadratic problems over polytope constraints, which are fast to solve. By constraining $y \in \mathcal{Y}^I$ plugging in \eqref{remarkableeq} in \ref{mc-BLP} and removing the already optimised $x$ variables, we obtain
\begin{equation*}
\begin{aligned}
    \min_{y}& \sum_{z=1}^{N_z}\alpha_z(\mathcal{Y}^I_z)y_z^2 + \left[\beta_z(\mathcal{Y}^I_z)-\alpha_z(\mathcal{Y}^I_z) Q_z^s(\mathcal{Y}^I_z)\right]y_z \\
    s.t. \hspace{1mm}&  M_p y \leq b_p \\
    & y \in \mathcal{Y}^I, \\
    & 1^Ty = d
\end{aligned}    ,
\tag{sub-CQP($\mathcal{Y}^I$)}\label{sub-CQP}
\end{equation*}
which is a much simpler problem to solve since $\mathcal{Y}^I$ is just a box constrained domain. The naive way of obtaining the global solution would then be to solve \ref{sub-CQP} for all possible $\mathcal{Y}^I$, and simply choose the best solution. However, their number number can be fairly large. In our case we observe about $11$ (including $(0,0)$) discontinuity points per zone, giving about $10^5=100,000$ combinations which is extremely large. This would in fact be much more computationally expensive than the BBTree approach, that also guarantees global convergence. However, a compromise can be obtained by iteratively updating the box location $\mathcal{Y}^I$ based on the previous result. The idea of the algorithm is as following:
\begin{enumerate}
    \item pick a feasible $y$, and solve \ref{sub-CQP} over the $\mathcal{Y}^I$ interval,
    \item if for any zone $z$ the corresponding optimal $y_z^*$ is on the boundary, then switch the bounds to be the other set of bounds $\mathcal{Y}^{I,2}_z$ that includes $y_z^*$. For example imagine that the initial domain for $y_z$ was $[y_z^{l_1}, y_z^{u_1}]$ and the next domain in increasing order is $[y_z^{l_2}, y_z^{u_2}]$ (with $y_z^{l_2}=y_z^{u_1}$). Then if the optimal solution with the $[y_z^{l_1}, y_z^{u_1}]$ bound gives $y_z^* = y_z^{u_1}$, the interval to solve over at the next iteration is updated to $[y_z^{l_2}, y_z^{u_2}]$. Interval updates for all zones can be performed simultaneously if necessary.
    \item Iterate until the optimal solution over some $\mathcal{Y}^{I,2}_z$ lies strictly inside the domain, or the change in $y$ between two consecutive (outer) iterations is zero.
\end{enumerate}
Note that in a strict sense, this approach does not guarantee global convergence, but it guarantees something close: global convergence over the domain $\mathcal{Y}^I$ in which the solution $y^*$ lies. This is a much stronger result than purely local convergence, although not as strong as purely global convergence. We refer to this type of optimality as sub-domain optimal, to distinguish it from the weaker counterpart of purely local optimality
\begin{algorithm}[h!]
\caption{Iteratively Bounded CQP(ib-CQP)}\label{alg:ib-CQP}
\begin{algorithmic}
\Require input $(m_i,a_i)$ for all $\forall i$, $(M_p,b_p)$, and demand $d$. 
\State Compute the Chebychev centre $y_c$ and the radius $r_c$ of the $(M_p,b_p)$ polytope. 
\State $y \gets y_c$ and set $\Delta y=1$
\While{$\Delta y>0$}
\State Get the domain $\mathcal{Y}^I$ that contains $y$ and the corresponding $\alpha$, $\beta$
\State Obtain $y'$ and $v'$ by solving \ref{sub-CQP} and compute $\delta y = \|y-y' \|_2$
\State $y \gets y'$, $v \gets v'$
\State \textbf{If} $y'$ not on boundary of $\mathcal{Y}^I$, \textbf{then} break
\State Update $\mathcal{Y}^I$ for all $z$ where $y_z$ is on the boundary
\EndWhile
\State use equation \eqref{x_i_get} to compute $x$ knowing $y$, $v$ and the active player set $\mathcal{P}^F \cup \mathcal{P}^M$
\State return $x$, $y$, $v$
\end{algorithmic}
\end{algorithm}
\par 
Note that since \ref{sub-CQP} makes no assumption with regards to the set of active players, this is obtained as part of the solution, instead of requiring any assumption or pre-computation as for the previously devised methods. This is a great advantage over the other proposed methods. The proposed approach is finally summarised in Algorithm \ref{alg:ib-CQP}, which further requires computing the optimal $x_i^*$ which is given by
\begin{equation}\label{x_i_get}
    x_i^* = \frac{y_{z_i}+\sum_{j \in \mathcal{J}_{z_i}}m_j^{-1}(a_j- a_i )}{m_i \sum_{j \in \mathcal{J}_{z_i}}m_j^{-1}},
\end{equation}
but one can alternatively solve the corresponding LP for fixed $y$.
\par
\begin{figure}[ht]
    \centering
    \includegraphics[width=7.9cm]{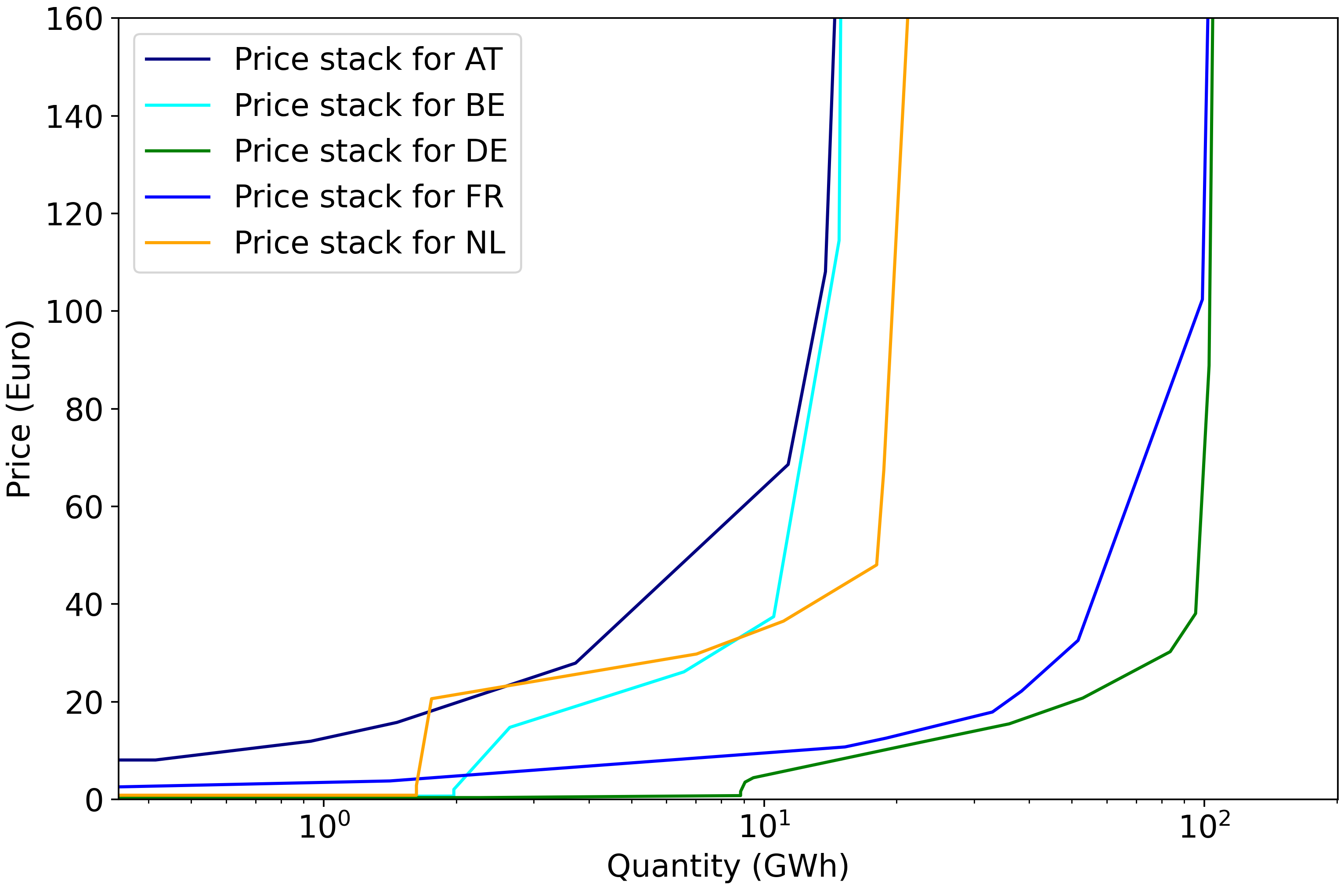}
    \caption{Obtaining the optimal price function: visualisation of the resulting (optimal) zonal stack curves.}
    \label{stacks}
\end{figure}

To help the reader get a better understanding of this approach, we show the obtained $v(y)$ for the real world case of CWE in Figure \ref{stacks} for each zone $z$.
Note that these are exactly analogous to the stack curve proposed in \cite{HowisonStack}, with one stack available per zone. The zonal network constraints force us to obtain optimal zonal stacks, in contrast to a single unified one for the whole market, to accommodate market decoupling. We can now see that the usual one dimensional stack model is a particular case of a cost minimisation model, when (partial) analytical solution is possible.
\section{Numerical results} 
In this section we first look at numerical results on synthetic data to get a better intuition about the behaviour of each market clearing solution approach we propose. We then then apply our models to the real world case of CWE data, and use the conclusions obtained on the synthetic data to aid us in the choice of the solution algorithm for the CM problem.
\subsection{Comparison of Numerical Approaches on Synthetic data}
We first look at computational times taken and quality of results for each algorithm on a simple CM market clearing problem. We then compute profit functions within a possible range of strategies. The reason for this is two-fold: (i) to provide multiple test problems to obtain more meaningful statistics and (ii) to understand the capability of each approach to obtain best response functions which could be then used under a Game-Theoretic framework. However, the focus remains on the properties of the approaches.
\par 
We consider a market with three zones, two players per zone, and the zonal demand vector is $d^T =\left[10, 6, 6\right]$. The production costs, capacity constraints and network constraints are defined by
\begin{equation}
    \begin{aligned}
    \hspace{0.25cm}&\hspace{0.0cm}c_0^T =\begin{bmatrix} 
 0.50 & 0.50 & 0.40 & 0.40 & 0.50 & 0.50 
\end{bmatrix},\\
\hspace{0.25cm}&\hspace{0cm}b_0^T = \begin{bmatrix} 
 1.50 & 1.30 & 0.80 & 0.63 & 0.20 & 0.40 
\end{bmatrix},\\
\hspace{0.25cm}&\hspace{0cm}Q_0^T = \begin{bmatrix} 
 4.00 & 5.50 & 4.00 & 4.00 & 3.00 & 4.00 
\end{bmatrix},\\
&M_p = \begin{bmatrix} 
 1.0 & -1.0 & 0.0 & \\
 1.0 & 0.0 & -1.0 & \\
 0.0 & 1.0 & -1.0 & \\
 1.0 & 0.0 & 0.0 & \\
 0.0 & 1.0 & 0.0 & \\
 0.0 & 0.0 & 1.0 & \\
 -1.0 & 0.0 & 0.0 & \\
 0.0 & -1.0 & 0.0 & \\
 0.0 & 0.0 & -1.0 & \\
\end{bmatrix},\hspace{1mm}
b_p = \begin{bmatrix} 
 5.0 \\ 5.0 \\ 2.5 \\ 12.0 \\ 8.0 \\ 8.0 \\ -8.0 \\ -4.0 \\ -4.0
\end{bmatrix}.
    \end{aligned}
\end{equation}
By using $m_0 =c_0$ and $a_0 =2 b_0$ to define the asks submitted to the CM market clearing problem, we can solve \eqref{CM-bl-BLP} using the four approaches proposed. The results are summarised in Table  \ref{tab:7.1rezTables}.

\begin{table}[ht!]
\centering
    \begin{tabular}{|c|c|c|c|}
        \hline
         Method & Objective & Time & Convergence  \\
         & function & taken & indicator  \\
         \hline
        ieq-LP-ma & 78.56 & 0.12s & $\frac{\|x^i-x^{i-1} \|_2^2}{\|x^i\|_2}<10^{-6}$ \\ \hline
        ie-QP-wr & 77.463 & 0.21s & $\frac{\Delta z}{\|z_c \|_2} <2\cdot 10^{-3} $  \\ \hline
        BBTree & 77.463 & 37.3s & $\delta_T < 10^{-5}$ \\ \hline
         ib-CQP (I) & 77.512 & $\approx$0.005s & $\frac{\|y^i - y^{i-1} \|_2}{ \| y^i\|_2 } < 10^{-2}$ \\ \hline
         ib-CQP (II) & 77.463 & $\approx$0.01s & $\frac{\|y^i - y^{i-1} \|_2}{ \| y^i\|_2 }< 10^{-4}$ \\ \hline
    \end{tabular}
    \caption{Objective functions, times taken and convergence indicators for ieq-LP, ie-QP, BBTree, ib-CQP. The labels (I) and (II) for ib-CQP refer to CQP solution tolerances of $10^{-2}$ and $10^{-4}$ respectively.}
    \label{tab:7.1rezTables}   
\end{table}
It can be observed that ie-QP-wr and ib-CQP reach the same objective value as BBTree which can be confirmed to be the global minimum up to an optimality gap of only $10^{-5}$, which can be viewed as the true solution. However, ib-CQP is by far the fastest, since it leverages strong analytical results, while ie-QP-wr is still relatively fast, and much faster than BBTree. The extra computational time paid to solve via BBTree is the cost one pays for a global optimality certificate. Both ib-CQP and ie-QP-wr have (stronger than) local convergence guarantee, but these are not truly global. Further, reaching the true solution via ib-CQP required a sufficiently small tolerance when solving each Convex Quadratic Program. When this tolerance is chosen as $10^{-2}$, the solution is twice as fast, but at the cost of accuracy degradation. Finally, while ieq-LP-ma(II) is faster than ie-QP-wr, at the cost of solution degradation, it is still slower than ib-CQP. The reason for this is that ieq-LP-ma does not leverage the problem structure at maximum, and does not offer any convergence guarantees. As a result, when solving via ieq-LP-ma, a large fraction of time is spent ensuring that the distance between two iterations becomes very small. This confirms that ieq-LP-ma is best used as a way to quickly compute an upper bound for nodes in BBTree, rather than simply by itself.
\par
We now compute approximate profit functions for each player over the range $m_i \in [0.5c_i, c_i]$, and $N_{pts}=20$ discrete equally spaced strategy points for player $i$. The corresponding value of $a_i$ is obtained as $a_i = \nu_i - \mu_i m_i$, with $\nu^T = (3.4, \hspace{0.4mm}3.4, \hspace{0.4mm}1.815, \hspace{0.4mm}1.815, \hspace{0.4mm}1.55, \hspace{0.4mm}1.55)$, and $\mu^T = (1.89, \hspace{0.4mm},2.1, \hspace{0.4mm}1.27, \hspace{0.4mm}1.38, \hspace{0.4mm}1.35, \hspace{0.4mm}1.15)$. For all other players we consider bidding structures, given by $(m^I,a^I)=(1.35c_0,1.02b_0)$, $(m^{II},a^{II})=(2c_0,b_0)$.
\begin{figure*}[ht!]
\centering
\includegraphics[width=1.69\columnwidth]{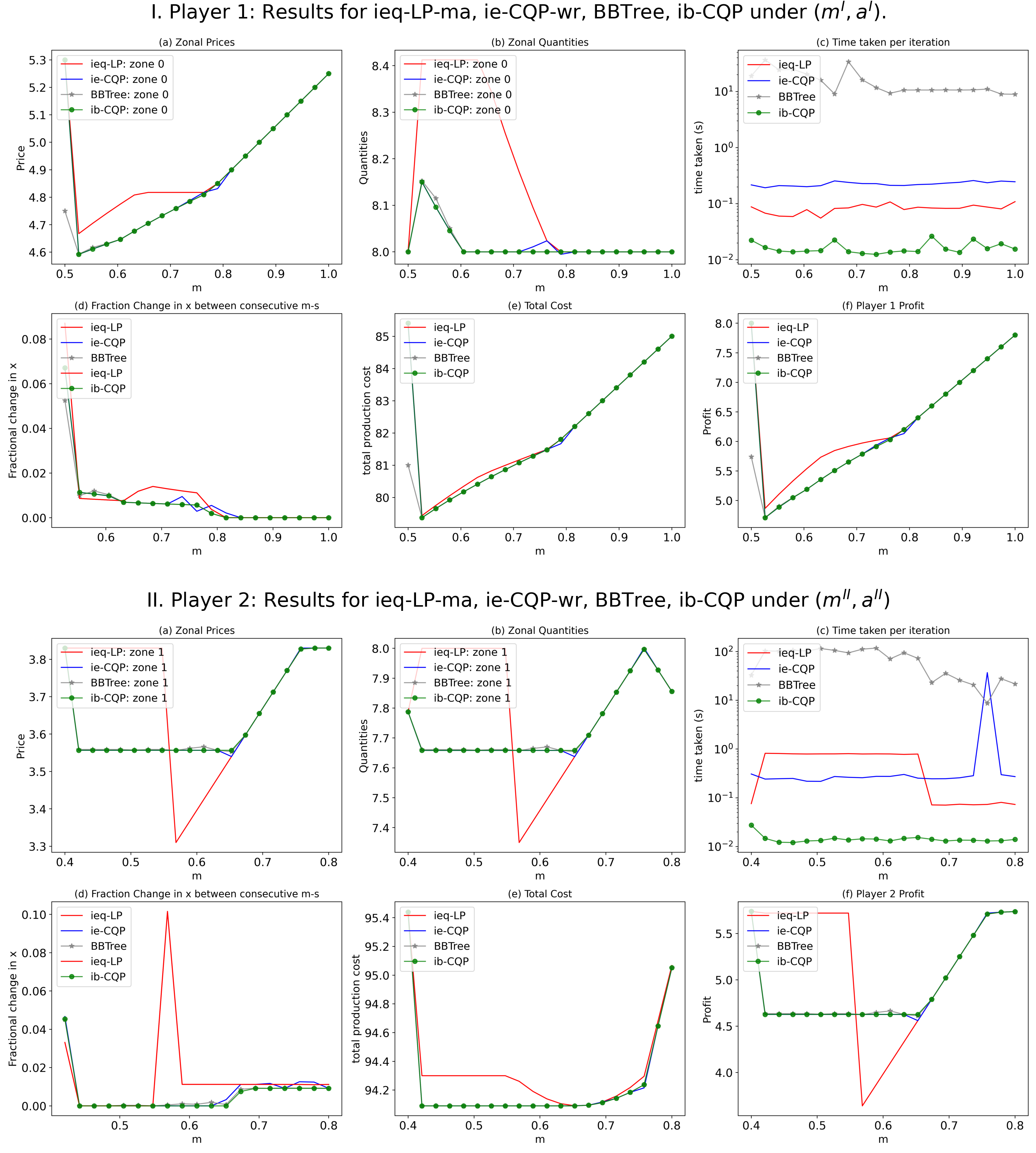}
\caption{Representative results for proposed numerical approaches on synthetic data.}
\label{fsynth}
\end{figure*}
We show representative results for the four approaches in Figure \ref{fsynth}. In this case we can take the results given by BBTree as the true solution. We can see that both ib-CQP and ie-QP-wr closely replicate the solution given by BBTree, but ib-CQP generally performs best as it is much faster. This is also confirmed by Table \ref{tab:7.2rezTables}, where we summarise the total computational time for computing the profit functions over the specified intervals. We can see that the cost for guaranteeing a global solution via BBTree can be very high, while ib-CQP can generally obtain the same solution, but it offers a weaker convergence guarantee. However, this  convergence guarantee is still very strong: global convergence on the corresponding subdomain as explained in Sub-section \ref{sec:ib-CQP}. For this reason, we chose ib-CQP to compute solutions to the CM problem when looking at the real world CWE case study.
\begin{table}[h]
\centering
    \begin{tabular}{|c|c|c|c|c|}
        \hline
        & ieq-LP & ie-QP & BBTree& ib-CQP\\
        \hline
        \hline
        Player 0 &1.42s &4.5s&112s&0.33s\\
        \hline
        Player 1& 4.45s &4.32s&154s&0.25s\\
        \hline
        Player 2& 1.44s &25.1s&456s&0.51s\\
        \hline
        Player 3 &1.51s& 5.4s& 611s &0.61s\\
        \hline
        Player 4 & 0.98s&4.2s&556s&0.78\\
        \hline 
        Player 5 &1.4s&5.5s&63s&0.31s\\
        \hline
    \end{tabular}
    \caption{Time taken to compute profit functions for each player, with $N_{pts}=20$ discretisation points on $\mathcal{B}_{LB}(.)$}
    \label{tab:7.2rezTables}   
\end{table}
\begin{figure*}[ht]
    \centering
    \includegraphics[width =15.5cm]{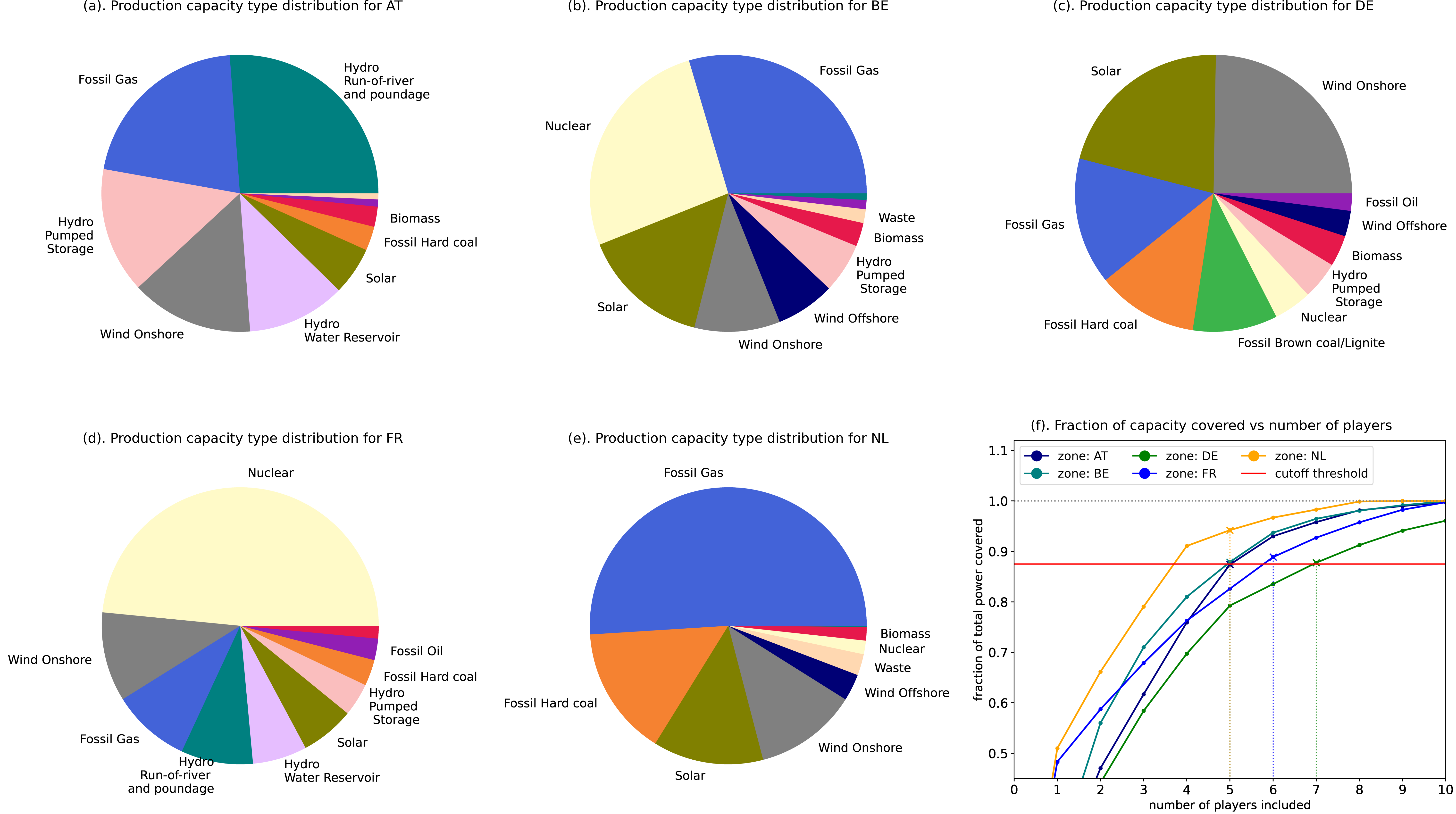}
    \caption{Capacity per production types per country, and illustration of player choice. Sub-figures (a)-(e) show the largest 10 production types per country. Sub-figure (f) illustrates the thresholding process.}
    \label{fig:caps}
\end{figure*}
\newpage
\subsection{Real world case study: Central Western European spot market}\label{sec:numres} 
Our approach was developed with the Central Western European Day Ahead Auction in mind, and therefore naturally, we are interested in applying this methodology to CWE DAA data to obtain forecasts of market outcome and prices based on fundamental inputs. To evaluate the suitability of our proposed model, the obtained prices are then be compared to the actual real world prices, referred to as \textit{target prices} from here on. Most computations in this section are performed using \textit{Descartes Labs}' computing services \citep{desclab}.
\subsubsection{Input data}
We run our models on the real world case of CWE, for each hour (indexed by $t$) independently, for $2712$ hours starting at first hour of 2 Jan 2019. This requires temporal time series for the demand series $d(t)$, network constraints $\left\{M_p(t),b_p(t)\right\}$ and player's characteristics $\left(c_0(t),b_0(t),Q_0(t)\right)$. The demand time series can be obtained from \cite{ENTSOEData}, and we use the day ahead forecasted values as inputs. 
\par 
The time series corresponding to the network constraints, $\left\{M_p(t),b_p(t)\right\}_{t \in \mathcal{T}}$ is obtained by first obtaining the full set of historical PTDFs and RAMs as described in \cite{PuiuHauser2021}. Let us denote the set of line-scenario combinations \cite{PuiuHauser2021} that are tracked by the operator as $\mathcal{LS}$. The concatenation of all these constraints at time $t$ is then given by 
\begin{equation}
    r^{\mathcal{LS}} \leq PTDF_t^{\mathcal{LS}}(y(t) - d(t)) \leq R_t^{\mathcal{LS}},
\end{equation}
and therefore the constraint set is 
\begin{equation}\label{polytopeformation}
    M_p(t) = \begin{bmatrix} - PTDF^{\mathcal{LS}}_t \\ +PTDF^{\mathcal{LS}}_t \end{bmatrix}; \hspace{1mm} b_p(t) = \begin{bmatrix} - r^{\mathcal{LS}}_t  - PTDF^{\mathcal{LS}}_td(t)\\ +R^{\mathcal{LS}}_t + PTDF^{\mathcal{LS}}_td(t) \end{bmatrix},
\end{equation}
but we also add to this box constraints to ensure that $0.4d_z(t) \leq y_z(t) \leq 1.6 d_z(t)$ to guarantee that the feasible domain is a polytope.
\par 
We are now left with obtaining producers' characteristics at each time $\left(c_0(t),b_0(t),Q_0(t)\right)$. Note that $Q_0(t)$ may change due to power-plant outages, but this effect may be less important than the change in costs, that may be very large and rapid, at least for power plants based on fossil fuels such as natural gas, coal or oil.
\par 
 We first have to define the player set. For simplicity, for each country, we define each player to represent all power plants for a specific type of production (e.g. one player for all Gas plants in Austria), and consider all main production types as per the ENTSOE data on \textit{Installed Capacity per Production Type}. This would give us around $11$ players per country, and thus $55$ players in total. However, some of these players would have very small capacity for some countries, and thus essentially do not affect the market outcome, but significantly increase the computational cost. Since we are required to solve for thousand of time indexes and the set of network constraints is composed of more than $600$ rows, the total computational cost can grow very large. Therefore, we take a very parsimonious approach and restrict the player set, in an attempt to minimise computational effort while retaining the dominant fundamental factors. To this end, for each zone, we retain the top $n_z$ players that cover at least $\theta = 88\%$ of the total installed capacity in the country, but require that $n_z\geq 5$,  to retain most of the diversity of production types as well. 
 \par
In other words, if $\phi_z(n_z)$ is the fraction of total capacity covered by the largest $n_z$ players, then for each $z$ we choose $n_z^*= \min\{n_z \hspace{0.1mm}: \hspace{1mm} \phi_z(n_z)\geq 0.88, n_z \geq 5, \hspace{1mm}, \hspace{1mm} n_z \in \mathbb{N}\}$. Figure \ref{fig:caps} (a)-(e) show the top $10$ production types for each country, while sub-figure \ref{fig:caps}.(f) summarises the approach for deciding the number of players. By using the numbers shown in this sub-plot, the reader can obtain exactly the types of production considered for each country, with the mentioning that On- and Off-shore wind, Hydro Pumped Storage and Hydro Water Reservoir, as well as Hard Coal and Lignite types are combined together in one player where possible, due to their similarity of cost and actual capacity structures. In this way we get 5 players for the AT, BE, NL zones, 6 for the FR zone, and 7 for the DE zone, and thus the total number of players is 28, about half of the initial number.
\par
With the player sets well defined, we can now proceed to gathering the cost data for each player. Note that in principle, for each player, we require a time series $(c_i(t), b_i(t))$. For the final series used as inputs, the same type in different zones may have the same or different cost values depending on the properties of the type and available data. We construct the players' cost per type, and the types to be considered are: (i) Fossil Gas, (ii) Coal, (iii) Nuclear, (iv) Solar, (v) Wind, (vi) Hydro Pumped Storage (vii) Hydro Water Reservoir, (viii) Hydro Run-of-river and poundage. For simplicity, for each player, the two cost parameters are obtained by extrapolating from a representative value indicating the production cost per MWh, for each time index $t$. This could (but does not have to) be the corresponding fuel cost for one MWh of production. Let this cost for type $\theta$ in zone $z$ at time $t$ be $k_{\theta,z}(t)$. Note that $\theta,z$ define a player, and call it $j$. To obtain two parameters we then define the cost function for player $j$ as
\begin{equation}
    \lambda_{j,t}(x) := k_{\theta,z}(t)\left(1 - \frac{f_{\theta}}{1-f_{\theta}} n_{\theta}\right)+ n_{\theta}\frac{k_{\theta,z} }{(1-f_{\theta})Q_j}x = b_j+c_jx \,
\end{equation}
where $\lambda_{j,t}(x)$ is the price asked at quantity $x$, $f_{\theta}$ is the fraction of total capacity at which the price asked is exactly $k_{\theta,z}(t)$, and $n_{\theta}+1$ is the number of times the production cost is higher at full capacity when compared to the base average level, $ k_{\theta,z}(t)$. We take the cost to be $k_{\theta,z}(t)$ at $50\%$ capacity level, i.e. $f_{\theta} =0.5$ for all types for simplicity and $n_{\theta} \in [0,1]$, to guarantee that $b_j \geq 0$, with the exact value of $n_{\theta}$ depending on the fuel type. We are now left with obtaining $k_{\theta,z}(t)$ and $n_{\theta}$. Since it is hard to obtain $n_{\theta}$, we just assign sensible values based on judgement. In this context when $n_{\theta}$ is close to $1$ we say that the cost per MWh increases very fast with produced quantity, while a value close to $0$ represents slow cost per MWh increase. We now discuss how $k_{\theta,z}$ is obtained and the $n_{\theta}$ values are assigned for each type:
\begin{enumerate}
    \item For gas power plants we use the daily time series of spot gas prices at the Dutch Title Transfer Facility (TTF) hub, available at EEX, \cite{PowerNextData}. We take $n_{gas} = 0.2$ and the daily time series is expanded to hourly by assuming constant value over all hours within a day.
    \item For the coal prices, we use a close proxy of the daily spot coal price data available at Argus \cite{argus}. We pad the weekends (when trading is closed) with the Friday value. The prices are stated in $\$/ton$ and we convert this to $\text{\officialeuro}/MWh$ by using the fact that $0.55tons$ of coal generate $1MWh$ on average, and a $\$/\text{\officialeuro} = 0.89$, thus requiring a scale-up of the value by multiplying by $1.62$. We choose $n_{coal} =0.4 $.
    \item Nuclear power plants' production cost is taken as the average cost per MWh, quoted at about $31\$$, or about $27.6\text{\officialeuro}$ using the same conversion rate. However, because nuclear power plants cannot easily switch on and off, we assume that they are prepared to sell at a loss of $50\%$ and thus their switch-off cost is $k_{nuclear,z}(t)=13.8\text{\officialeuro}/MWh$ $\forall t$. Due to low production flexibility we choose $n_{nuclear} =0.8$.
    \item For Wind and Solar energy, we take a very low value of $k_{nuclear,z}(t)=0.5\text{\officialeuro}/MWh$ $\forall t$, and $n_.=0.05$ since increasing output is not expected to increase costs per MWh significantly. While the cost per MWh may seem very low in comparison to the usual quotes, we stick to this number due to the awareness that renewable energy influx was subsidised by the European Government for the year of 2019, meaning that the net cost was likely close to zero or even negative. To avoid modelling issues where Wind and Solar are just as reliable as nuclear, the capacity is given by the day-ahead daily forecast for wind and solar generation available at ENTSOE \cite{ENTSOEData}. Thus, for these types of players, while their cost is constant, their capacity changes rather erratically with time.
    \item Hydro Pumped Storage cost is obtained by taking for each day, the average base load electricity price (for the off-peak hours) over a running window including the past week. We assign $n_{hydro-s}=0.2$.
    \item Finally, the Hydro Run-of-river power plants generally have a levelised cost of energy (LCOE) between $2\$/MWh-$ and $7\$/MWh$. Taking an average value and converting it we obtain a cost of about $k_{hydro-r,z}=8.45\text{\officialeuro}/MWh$, $\forall t$. We choose $n_{hydro-r}=0.1$.
\end{enumerate}
The approach employed here for obtaining time series of the cost parameters for each player is admittedly very simplistic. We attempt to compensate for this by performing a simple calibration of costs to better represent prices in Sub-section \ref{modelcal}. Although simplistic, we use this approach to facilitate focusing on the formulation, analysis, and solution of the market clearing and game theoretic models. Nevertheless, we show that with minimal fitting (only 2 parameters per zone), we obtain results that capture well the main features of the actual prices time series, with train and test errors of only $17.9\%$ and $24.5\%$ on average, even for large test window (28 days or 672 hours). This reveals the robustness of our fundamental model and suggests that the reconstructed network constraints (from \cite{PuiuHauser2021}) offer a satisfactory representation of the true constraints, at least for this application.
\subsubsection{Model Calibration}\label{modelcal}
The calibration approach is very simple and is constrained to adjusting $(c,b)$ values only. This is to minimise computational cost but also the number of fitted parameters, as in principle, our fundamental model should be representative of the market without requiring any fitting, if the cost structure and other inputs are descriptive. Thus, we use this mostly as a simple way to improve our guess on the cost structure. 
\par 
Since production costs may change we can still consider the cost to depend on the changing fuel prices and fit only the power plants' characteristics. To this end, for each zone $z$ we simply fit a scale parameter for each $c$ and $b$ variables
  \begin{equation}
      c^{final}_i = s_z^c c_i^{initial}; \hspace{1mm} b^{final}_i = s_z^b b_i^{initial} \hspace{1mm} \forall i \in \mathcal{I}_z,
  \end{equation}
giving a total of $10$ parameters to fit, and we initialise $s_z^.=1$, $\forall z$.
Regardless of whether CM or SWM market clearing is used, the model price can be expressed as $v_z=s_z^cc_{k_z}x_{k_z}+b_{k_z}$ for each zone $z$, where $k_z$ is the player with the highest marginal price in the zone and $x_{k_z}>0$. The advantage of defining the cost scales is that $\frac{\partial v_z}{\partial s_z^.}$ can be easily computed, regardless of the fact that $k_z$ changes. The optimisation objective is then
\begin{equation}
    F(s) = \frac{1}{N_T}\sum_{t=1}^{N_T}\sum_{z}(v_{z,t} - P_{z,t})^2,
\end{equation}
where $v_{z,t}$  and $P_{z,t}$ are the model and target prices respectively, at zone $z$ for time $t$. Note that the scale parameters $s = (s^c,s^b)$ are not dependent on time, and thus have a very strongly over-determined Nonlinear Least Squares Problem. Let
\begin{equation}
    F_t(s) = \sum_{z}(v_{z,t} - P_{z,t})^2.
\end{equation}
We then have that
\begin{equation}\label{gradrefeq1233}
    \frac{\partial F_t}{\partial s_z^c} = 2(v_{z,t} - P_{z,t})c_{k_z}x_{k_z}; \hspace{2mm} \frac{\partial F_t}{\partial s_z^b} = 2(v_{z,t} - P_{z,t})b_{k_z},
\end{equation}
which by summation, and collecting for all zones $z$, gives $\nabla_{s}F(s)$. The second derivatives can be similarly computed as 
\begin{equation}
    \frac{\partial^2 F_t}{\partial^2 s_z^c} = c_{k_z}^2x_{k_z}^2; \hspace{4mm} \frac{\partial^2 F_t}{\partial^2 s_z^b} = b_{k_z}^2; \hspace{4mm} \frac{\partial^2 F_t}{\partial s_z^b\partial s_z^c} = b_{k_z}c_{k_z}x_{k_z},
\end{equation}
with all other entries in the $\nabla^2_{s} F$ being 0. Both Gradient Descent and Newton's Method with line-search as globalisation method were tried, and Gradient Descent appears to be able to obtain lower values of the objective function value. Note that these scales can be obtained by fitting either using the SWM or the CM models.
\subsubsection{Performance and comparison of CM and SWM}
We consider the case of truthful bids (TB), i.e. $m=c_0$ and $a=b_0$. We perform three types of runs that are of interest for the focus of this subsection, on both $\mathcal{T}_{train}=\{1,2,...,2016\}$ and $\mathcal{T}_{test}=\{2017,2018,...,2712\}$: (i) we run SWM with the SWM-fit parameters and call these results SWM (SWM-fit), (ii) we run CM with the SWM-fit parameters and call these results CM (SWM-fit)
, and (iii) we run CM with the CM-fit parameters and call these results CM (CM-fit). We use (i) and (iii) to compare which model better represents the observed price data, while the comparison between (i) and (ii) reveals the potential benefits of Cost Minimisation over Social Welfare Maximisation under identical circumstances. We compute the average error as $e_z(\mathcal{T}) = \frac{1}{|\mathcal{T}|}\sum_{t \in \mathcal{T}}|v_{z,t}^{model}-P_{z,t}|$, and the relative average error as $\overline{e}_z(\mathcal{T})= e_z/ \mu_{t \in \mathcal{T}}(P_{z,t})$, where $\mu_{t \in \mathcal{T}}(P_{z,t})$ represents the mean of $P_{z,t}$ over the $\mathcal{T}$ time indexes set.
\par 
\begin{table}[ht]
\begin{center}
\begin{tabular}{|c|c|c|c| }
 \hline
 & SWM & CM  &CM \\
 & (SWM-fit)& (SWM-fit) &(CM-fit)\\
 \hline
  AT &$20.3\%$ & $27.4\%$ & $36.0\%$ \\ 
  \hline
  BE & $14.2\%$ &$38.8\%$ &$20.5\%$ \\ 
  \hline
  DE &  $26.1\%$&$35.6\%$& $32.3\%$\\
  \hline
  FR &$14.5\%$&$15.5\%$&$18.4\%$\\ 
  \hline
  NL &$13.6\%$&$32.8\%$&$21.4\%$\\
  \hline
  Aggregate & $17.9\%$&  $30.2\%$ &$26.3\%$ \\
 \hline

\end{tabular}
\end{center}
\caption{\label{Tab:Train_TB} Results for $\overline{e}_z(\mathcal{T}_{train})$ for each zone and aggregate.}
\end{table}

\begin{table}[ht]
\begin{center}
\begin{tabular}{|c|c|c|c| }
 \hline
 & SWM & CM  &CM \\
 & (SWM-fit)& (SWM-fit) &(CM-fit)\\
 \hline
  AT &$28.8\%$ & $33.1\%$ & $33.3\%$ \\ 
  \hline
  BE & $22.8\%$ &$40.8\%$ &$22.3\%$ \\ 
  \hline
  DE &  $30.5\%$&$49.7\%$& $33.3\%$\\
  \hline
  FR &$19.6\%$&$31.6\%$&$17.4\%$\\ 
  \hline
  NL &$18.5\%$&$36.1\%$&$20.9\%$\\
  \hline
  Aggregate & $24.6\%$&  $38.2\%$ &$26.5\%$ \\
 \hline
\end{tabular}
\end{center}
\caption{\label{Tab:Test_TB} Results for $\overline{e}_z(\mathcal{T}_{test})$ for each zone and aggregate.}
\end{table}
\par 
As it can be observed from Tables \ref{Tab:Train_TB} and \ref{Tab:Test_TB}, SWM (SWM-fit) model is more representative for the market when compared to CM (CM-fit), for both the training and the test set. This is to be expected, since we know that the market clearing mechanism used in practice is a modified version of the SWM model we use.
\par 
We also observe that CM (SWM-fit) generally gives larger error because he SWM-fit parameters are used, instead of optimal parameters for the CM model. We can also observe that the test error is generally larger for the SWM, raising the question of over-fitting. However, the number of parameters is very small in relation to the complexity of the problem, and the SWM (SWM-fit) model still has the best generalisation error. 
\begin{figure}[ht!]
    \centering
    \includegraphics[width =7.4cm]{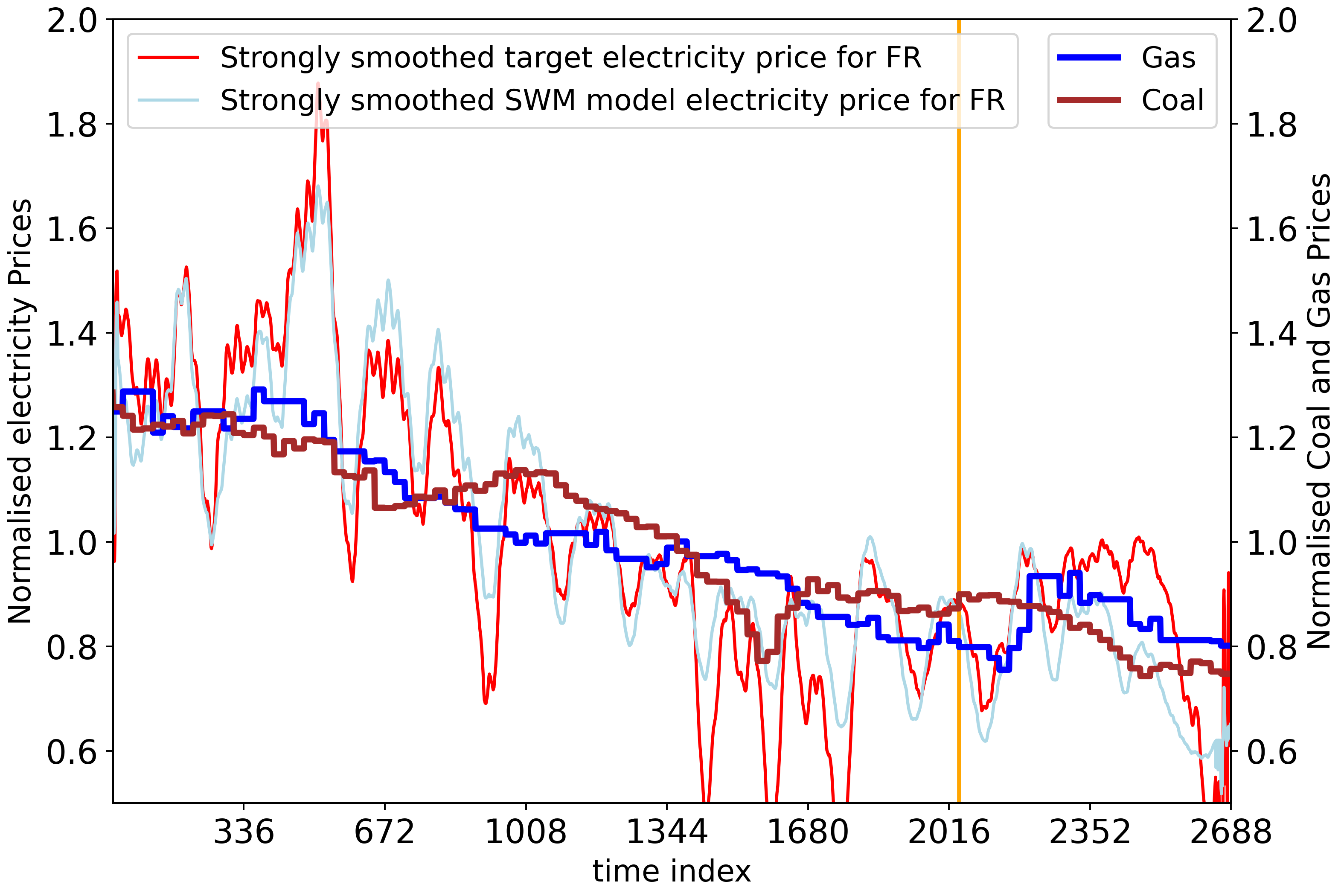}
    \caption{Trends Comparison: electricity, gas and coal price vs SWM-TB model prediction, for the $\mathcal{T}_{train}$ (left of orange line), and $\mathcal{T}_{test}$  (right of orange line). Electricity prices are smoothed by taking an exponentially weighted average over 49 hours centred in the point of interest. Only FR electricity prices are shown for better visualisation.}
    \label{train_test_underlying}
\end{figure}
Further, as can be seen in Figure \ref{train_test_underlying}, the test set is quite challenging. First we observe that on the second part of the test set, while gas and coal prices steadily decrease, on average, the electricity prices remain elevated and then suddenly drop. As a thought exercise, if we assume that all the other inputs remain constant, reduction in coal and gas prices will result (under TB) in our model predicting a reduction in electricity prices (as observed), especially since coal and gas power plants are often on the margin. Thus, this may increase the discrepancy between our prediction and the price target data. This behaviour of (smoothed) electricity prices suggests that producers may price their fuel costs in a rather conservative way, by taking higher and/or older prices as reference.

We next look at detailed results of the SWM-TB (SWM-fit) and (CM-TB (SWM-fit) models for the price time series and total costs, and compare this to the observed data. We only focus on the first $672$ and $336$ hours of the train and test sets respectively, for better visualisation of details. 
\begin{figure*}[ht!]
    \centering
    \includegraphics[width = 15cm]{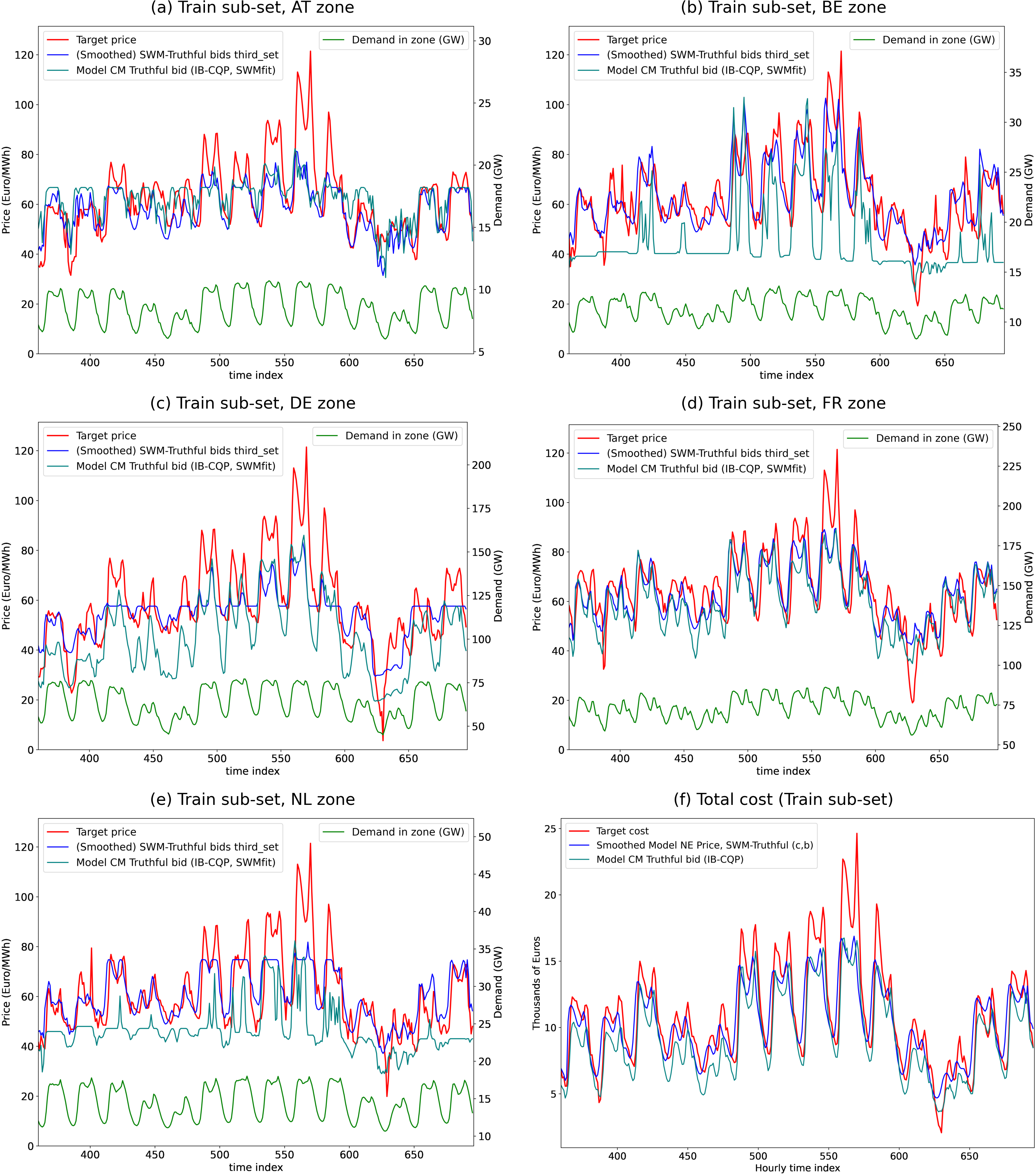}
    \caption{Train set: SWM-TB and CM-TB model results comparison with target prices, with indicative demand series.}
    \label{fig:R1.3.Train}
\end{figure*}
From Figure \ref{fig:R1.3.Train} we observe that the CM-TB model generally gives lower prices than the SWM-TB one, and that the total cost is always lower. The latter is expected, since by definition the CM optimisation problem is minimising this cost, while the SWM problem minimises some form of apparent cost. Further, CM gives much flatter price profiles, suggesting that when compared to SWM it reduces the impact that a single producer has on the final zonal price. 
We can further observe that the CM model price closely follows the SWM model price for France, revealing robustness of prices in this zone, with respect to the market clearing method, especially when compared to Belgium and the Netherlands, where the price reductions obtained via CM market clearing are very large. This reveals high price sensitivity of the Belgian and Dutch zones with respect to the market clearing rule. Since the Netherlands rely heavily on gas power plants, it also indicates that using the CM model (instead of SWM) could reduce the price impact of this reliance. One can also observe that CM often gives higher prices than SWM for the Austrian zone. This is not an inconsistency, as the cost minimisation problem guarantees that the total cost of electricity procurement is minimised, not the individual zonal costs or prices. It thus appears that if switching to Cost Minimisation, Austria may pay the price for obtaining cheaper electricity for the whole market on average. This is perhaps due to the large hydro storage capacity in this zone, which is optimally used by the clearing mechanism to minimise the total cost, instead of the zonal price in Austria, at least if producers bid truthfully. We note that the approach used for valuation of hydro stored power is likely to significantly impact this conclusion.
 \begin{figure*}[ht!]
    \centering
    \includegraphics[width = 15cm]{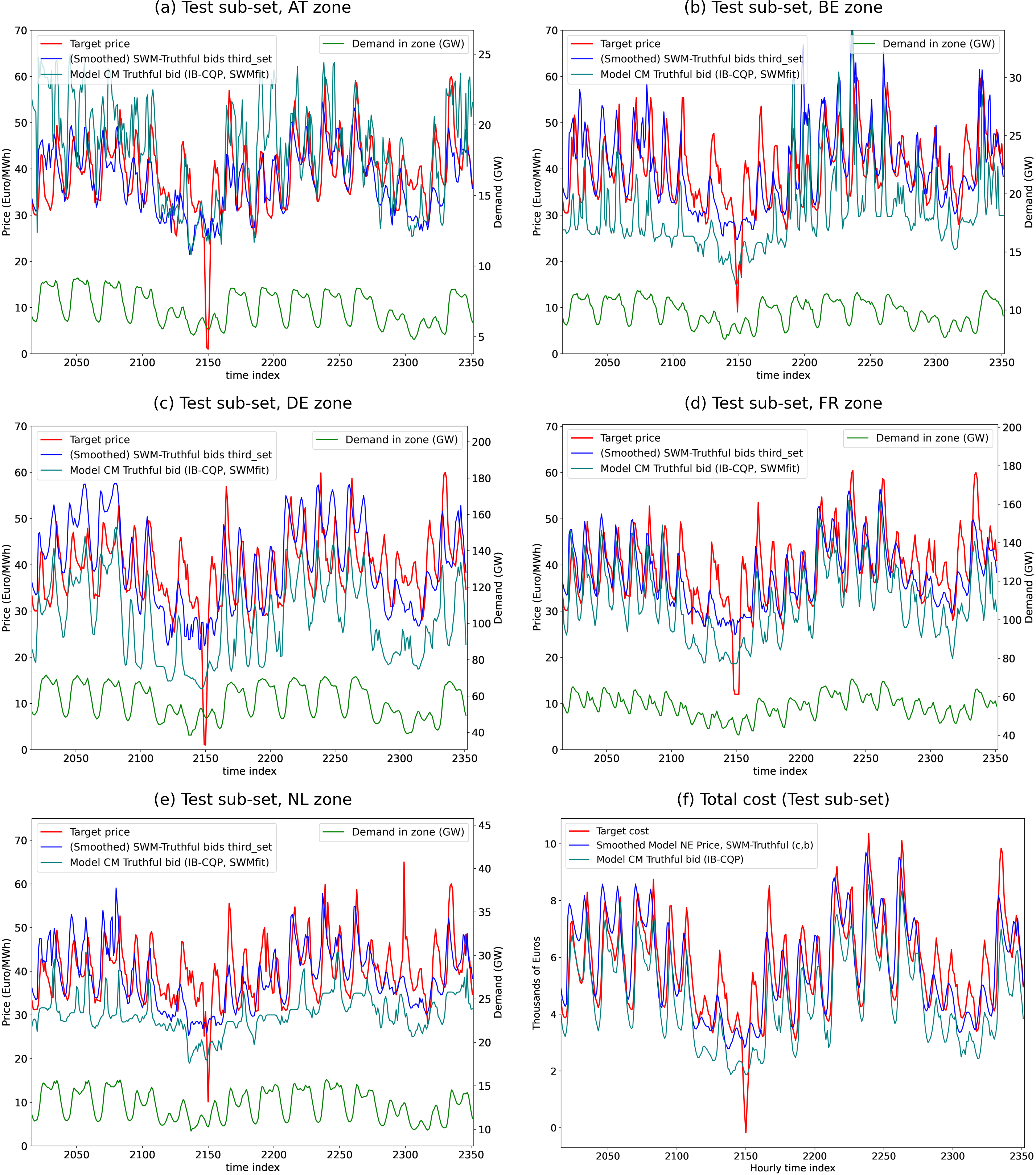}
    \caption{Test set: SWM-TB and CM-TB model results comparison with target prices, with indicative demand series.}
    \label{fig:R1.3.Test}
\end{figure*}
\par
 By now focusing on the comparison of SWM-TB with the target prices, we observe that often the SWM-TB model under-predicts the real price, despite the fact that no demand spike is recorded. The price spikes could therefore be a consequence of supply shortage or strongly constrained network. While we assume constant supply, variation in network constraints is included in the model. Assuming accurate network constraints, this suggests that players may bid strategically at times to maximise profits, despite the fact that they may often bid truthfully. Note that in practice it may be difficult to distinguish between the case of capacity reduction due to outages and capacity reduction as a strategic tool for increasing profit.
 \par
 We finally observe that our results are worst for the German zone, and this could be due to the very complex nature of this region, including a very diversified portfolio of power plants, large grid with multiple operators, and large renewable energy but also coal plants capacity. The nature of renewable production via wind and solar is very unpredictable, while the extent to which coal power plants are used may not be a function purely of market price, but also of environmental conscience, and in particular even when considering emissions fees, one may choose not to use profitable coal plants, if other sources are available.  Further error could be introduced by the modelling of participants, as the number we consider for Germany is only seven, while in reality there is a much larger number of market participants.
  \par From Figure \ref{fig:R1.3.Test} we can observe that SWM-TB generally performs well on the test set, that the CM-TB (SWM-fit) gives more frequent spikes than for the train set, suggesting that this set might have more erratic features. 
  \par
\begin{figure}[ht!]
    \centering
    \includegraphics[width = 7.5cm]{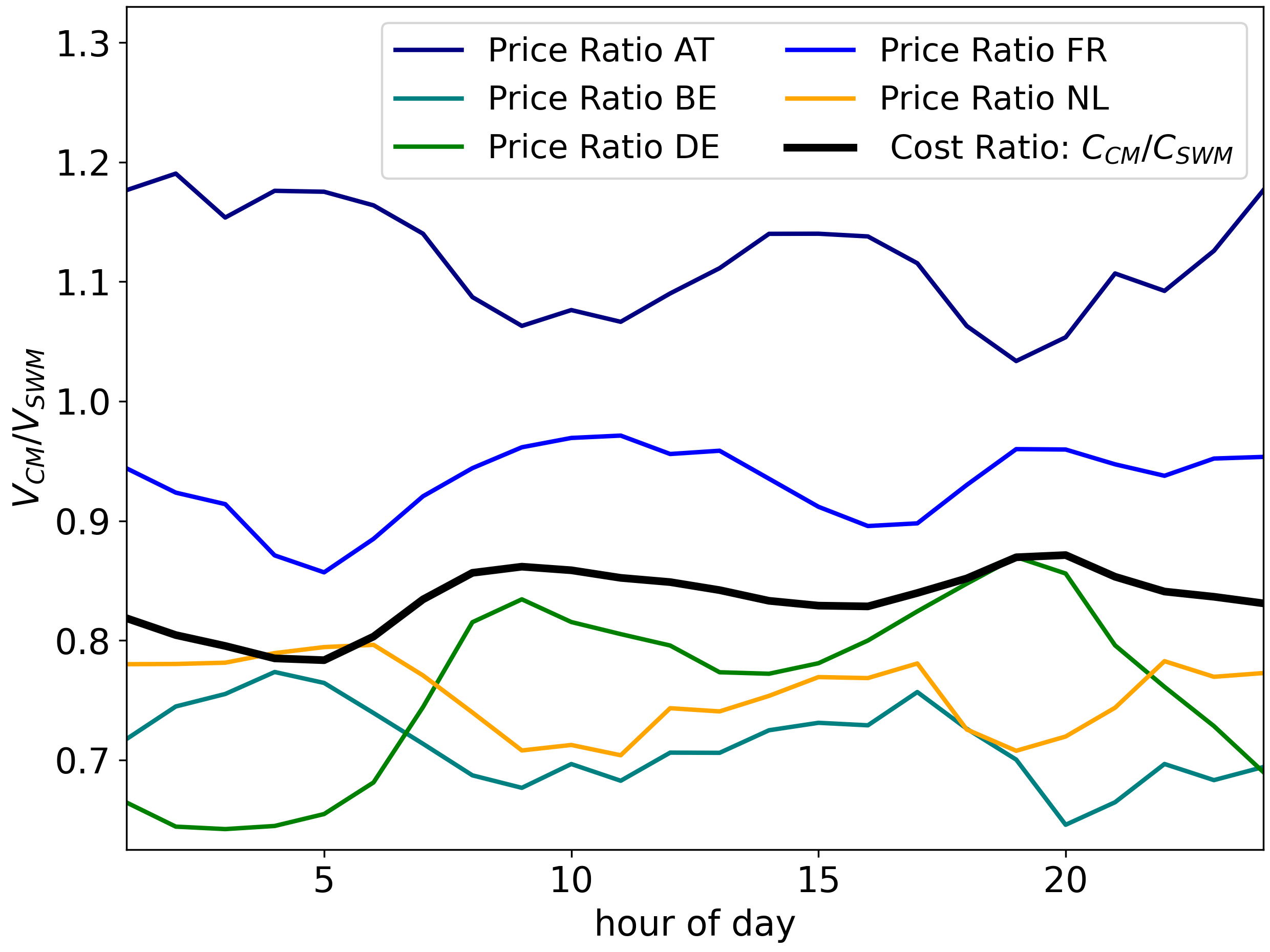}
    \caption{Price under CM to price under SWM ratio and the the total cost ratio.}
    \label{fig:R3.2.}
\end{figure}

  We finally look at the average daily price profile for each zone, and the corresponding ratio of the CM-TB (SWM-fit) to SWM-TB (SWM-fit), as per Figure \ref{fig:R3.2.}.
First, note that when considering a country, under inflexible demand, the price and cost ratio for that country is the same. We observe that indeed the CM model decreases the average total cost, and that the average daily prices for all zones except Austria are also reduced, for each hour. We can observe that Belgium, France and the Netherlands obtain the most drastic price reduction at hours of peak demand. This is because both France and Belgium have rather large production (nuclear) capacity at low prices. When compared to SWM, CM allocates power plant production levels much more effectively, and the effect of this allocation is more pronounced when demand levels are high. This is because the zonal price is set by the power plant producing at the highest cost, and thus reducing the allocation to expensive plants even by a small amount can have a large impact on the price. It is interesting to observe that the Netherlands also benefits of this effect. This could be due to cross-border flows from Belgium and France, as well as coal power plants which may produce at lower price levels than gas plants. 
\par
Thus, we conclude that our SWM-TB (SWM-fit) model is generally most representative for the price data, while using a CM market clearing model may generally give lower prices, the corresponding total electricity procurement cost is guaranteed to be lower. Further, the CM model may offer less sensitive prices with respect to underlying changes such as production costs, demand levels or even network constraints. However, it appears that there exist time intervals where the SWM-TB model is rather poor at explaining the actual price, and this occurs when there are large downward or upward price spikes. While the former probably results due to a large, unexpected influx of renewable power generation, the large upward spikes are likely due to strategic bidding, which we investigate in great detail in another work.
\section{Conclusion}
We proposed a Social Welfare Maximisation (SWM) and a Cost Minimisation (CM) market clearing model considering network constraints, that are directly applicable to European spot electricity markets considering Flow-Based Market Coupling and Zonal Marginal Pricing, and to the Central Western European spot electricity market in particular. We show how the SWM problem can be solved for efficiently and provide numerical algorithms to solve for the more challenging CM problem, with different computational speed and convergence guarantee trade-offs. Some insight into these trade-offs is provided via numerical experiments on synthetic data. We obtain strong (but partial) analytic results for the CM problem, and show how analytically computing what we call \textit{zonal stack curves} greatly simplifies the computational work required to numerically obtain an optimal solution of the CM problem. 
\par 
We then perform numerical simulations on the real world case of Central Western European Day Ahead Electricity market, to study the performance of both our SWM and CM models. We conclude that the SWM model represents the historical time series of prices well. By contrast with SWM, the CM model reduces producers' market power by generally resulting in reduced market prices, and always reducing the total procurement cost. From a policy perspective, this suggests that using a CM market clearing model is desirable if wholesale electricity prices are unacceptably high. This comes at increased problem complexity, but we provided efficient algorithms to resolve the market clearing problem.  
\par 
\newpage
\appendix

\section{Proofs of Theorems and Results }
\label{sec:sample:appendix}
\begin{nono-theorem}\label{th322.a}
Let the social welfare maximisation problem, its dual and the pricing problem be defined as above. Then, given inelastic demand, $\Pi_z^* = \max_{o \in \mathcal{O}_z}\{P_o^z: x_o^z>0 \}$ is the optimal solution to the optimisation problem given by \eqref{eq15} and \eqref{eq16}, and $\Pi^*$ does not yield any paradoxically accepted orders for the primal problem.
\begin{proof}
Let $P_O^z$ and $P_B^z$ denote the maximum and minimum prices corresponding to accepted offers and bids respectively. Then, we have that by definition $P_O^z=\max_{o \in \mathcal{O}_z}\{P_o^z: x_o^z>0 \}$ and $P_B^z=\min_{b \in \mathcal{B}_z}\{P_B^z: x_b^z>0 \}$. But since the demand is inelastic, $P_B^z := P_O^z$. Thus, by definition $P_O^z = \Pi_z$ and so the objective function \eqref{eq15} has a value of zero, which is the minimum attainable value since this is a sum of squares. 
\par
We next show that constraint \eqref{eq16} is satisfied. Since the demand is inelastic, the objective function of the primal problem becomes:
\begin{equation}
        SW(x_.) = -\sum_{z \in \mathcal{Z}}   \sum_{b \in \mathcal{O}_z}Q_o^zP_o^zx_o^z,
\end{equation}
the flow feasibility constraint becomes
\begin{equation}
     M_p E_ox_o \leq b_p - M_pE_bx_b =:\overline{r},
\end{equation}
and the supply and demand balance:
\begin{equation}
    e_o^Tx_o=e_b^Tx_b,
\end{equation}
where $x_o=||_v(x_o^z)$ is the concatenation of all offers for all zones (and similar for $x_b$). Note that since the demand is inelastic, $x_b^z =1 $ for all $b$ and $z$ and thus these are not variables.
The lagrangian of the primal and dual is
\begin{equation}
\begin{aligned}
    L =& -SW(x_o) - \gamma^T(\overline{r}-M_pE_ox_o) \\&- y(e_b^Tx_b-e_o^Tx_o) - p^T(1 - x_o) -w^Tx_o.
\end{aligned}
\end{equation}
Taking the partial derivative w.r.t. every offer $i$ and setting to zero, we get
\begin{equation}\label{lagrder}
    \frac{\partial L}{\partial x_i}= -\frac{\partial SW}{\partial x_i} + \gamma^T(M_pE_o)_{:,i} - y Q_i^z+\tilde{p}_i= 0,
\end{equation}
where $\tilde{p}_i = p_i-w_i$. These are the optimality conditions for the social welfare maximisation problem. We note that $(AE_o)_{:,i}=A (E_o)_{:,i}$, and that by definition of $E_o$, $(E_o)_{z,i} = Q_i^z $ if $i \in \mathcal{O}_z$ and $(E_o)_{z,i}=0$ otherwise. There is exactly one such nonzero entry for every $i$ since an offer can only be placed in exactly one zone. Thus
\begin{equation}\label{Apart}
    \gamma^TM_p (E_o)_{:,i} = \gamma^T(M_p)_{:,z}Q_i^z.
\end{equation}
Finally, by differentiation of the objective function $SW$ we get
\begin{equation}\label{SWpart}
    -\frac{\partial SW}{\partial x_i} = Q_i^zP_i^z
\end{equation}
By virtue of \eqref{lagrder}, \eqref{Apart} and \eqref{SWpart} we get that
\begin{equation}
    Q_i^zP_i^z +\gamma^T(M_p)_{:,z}Q_i^z - y Q_i^z +\tilde{p}_i=0
\end{equation}
By simplifying $Q_i^z$ (since $Q_i^z>0$), we get 
\begin{equation}\label{Price-M-gamma}
    P_i^z = -\gamma^T (M_p)_{:,z} + y - \tilde{p}_i/Q_i^z
\end{equation}
and by taking the difference for the indices $i_z^* = \arg \max_{o \in \mathcal{O}_z}\{P_o^z: x_o^z>0\}$, we get that
\begin{equation}\label{Price-M-gamma2}
    P_{i_{z_1}^*}^{z_1} -P_{i_{z_1}^*}^{z_2} = \gamma^T((M_p)_{:,z_2}- (M_p)_{:,z_1}) + \tilde{p}_{{i_{z_1}^*}}/Q_{i_{z_1}^*}^{z_1} - \tilde{p}_{{i_{z_2}^*}}/Q_{i_{z_2}^*}^{z_2}
\end{equation}
We next prove that for the maximal indices $ \tilde{p}_{{i_{z_1}^*}}, \tilde{p}_{{i_{z_2}^*}} = 0$. For any non-trivial quantities traded, it must be the case that $ w_{{i_{z_1}^*}}, w_{{i_{z_2}^*}} = 0$, since the corresponding quantities must be nonzero. We are left to next prove that for the maximal indices $ p_{{i_{z_1}^*}}, p_{{i_{z_2}^*}} = 0$ for any $z_k$.
\par
Suppose that this is not true, and thus $p_{{i_{z_1}^*}}>0$, and so by complementarity slackness $x_{i_{z_1}^*}^{z_1}=1$. Consider the two options:
\begin{enumerate}
    \item assume there exists an $x_i^{z_1} \in (0,1)$. Then this must correspond to the maximal price, since the optimality can be obtained via a greedy solution. However, we have that $x_{i_{z_1}^*}^{z_1}=1$, implying that  $i_{z_1}^* \neq \arg \max_{o \in \mathcal{O}_{z_1}}\{P_o^{z_1}: x_o^{z_1}>0\}$. We have reached a contradiction.
    \item The only other alternative is that $x_i^{z_1} \in \{0,1 \}$ for all $i \in \mathcal{O}_{z_1}$. But then consider $x_{i_{z_1}^*}^{z_1}\leq r_{i_{z_1}^*}^{z_1}=1$. We have that \begin{equation*}
        \frac{\partial L}{\partial r_{i_{z_1}^*}^{z_1}} = p_{i_{z_1}^*}.
    \end{equation*}
    Increasing $r_{i_{z_1}^*}^{z_1}$ for $i_{z_1}^*$ will not change the solution since $x_i^{z_1}=1$ for $\forall i \in \mathcal{O}_{z_1}\setminus \{i^*_{z_1}\}$, and $i^*_{z_1}$ is the player with the maximum price. Thus $0= \frac{\partial L}{\partial r_{i_{z_1}^*}^{z_1}} = p_{i_{z_1}^*}$.
\end{enumerate} 
Thus,
\begin{equation}\label{Price-M-gamma2_trimmed22}
    P_{i_{z_1}^*}^{z_1} -P_{i_{z_1}^*}^{z_2} = \gamma^T((M_p)_{:,z_2}- (M_p)_{:,z_1}).
\end{equation}
But this always implies \eqref{eq16} since this system is obtained by collecting all the equations \eqref{Price-M-gamma2_trimmed22} for all the possible combinations $\{z_1,z_2\}$. Thus $\Pi^*$ is feasible reaching minimum objective value: an optimal solution to the minimisation problem given by \eqref{eq15} and \eqref{eq16}. Since the objective is (strictly) convex, and the best possible value is bounded below zero, the zero objective value can be attained at most at one point. But $\Pi^*$ yields this value and thus it is the unique optimal solution.
\par 
Finally, we note that all bids are accepted by definition of the problem, and thus they cannot be paradoxically accepted. Further, all the offers accepted in the social welfare maximisation problem must have $P_i^z \leq \max \{P_i^z:x_o^z \}=:\Pi_z^*$ and thus they are below the unique zonal price, meaning that they will not be rejected after price determination.
\end{proof}
\end{nono-theorem}


\section*{Acknowledgements}
We would like to thank CGM London Power and Gas desk within Macquarie Group for partially funding this research, and providing computational resources via \textit{Descartes Labs}. \textit{Ioan Alexandru Puiu} would like to further thank \textit{Vincent Guffens} for his support, \textit{Descartes Labs} for extending the account, and \textit{Jeremy Malczyk} and \textit{Chris Moulton} for offering technical support with the Descartes Labs platform.
\bibliographystyle{elsarticle-num-names} 
\bibliography{cas-refs}





% \end{thebibliography}
\end{document}